\documentclass[aps,prx,twocolumn,amsmath,nofootinbib,amssymb,superscriptaddress]{revtex4-1}
\usepackage{graphicx}
\usepackage{soul}
\usepackage{verbatim}
\usepackage{esint}

\usepackage[colorlinks=true,citecolor=blue,linkcolor=magenta]{hyperref}
\usepackage{amsmath}
\usepackage{graphicx}
\usepackage{dsfont}
\usepackage{changepage}
\usepackage{fancyhdr}
\usepackage{amsthm, amssymb}
\usepackage{array}
\usepackage[usenames,dvipsnames]{color}

\def \be{\begin{equation}}
\def \ee{\end{equation}}
\def \ba{\begin{array}}
\def \ea{\end{array}}
\def \bea{\begin{eqnarray}}
\def \eea{\end{eqnarray}}
\def \nn{\nonumber}
\def \ve {\varepsilon}
\renewcommand{\vec}[1]{\boldsymbol{#1}}

\def \etal{{\it {et al. }}}

\def \W{{\Omega}}
\def \e{{\epsilon}}
\def \l{{\lambda}}

\def \a{{\alpha}}

\def \b{{\beta}}
\def \g{{\gamma}}

\def \d{{\delta}}
\def \w{{\omega}}
\def \s{{\sigma}}

\def \e{{\epsilon}}
\def \ve{{\varepsilon}}

\def \G{{\Gamma}}

\def \ba{\begin{align*}}
\def \ea{\end{align*}}

\newcounter{indice}

\def \mrm{\mathrm}

\def \bs{\boldsymbol}
\def \mc{\mathcal}

\begin{document}

\title{Superconductivity in dilute SrTiO$_3$: a review}
\author{Maria N. Gastiasoro}
\affiliation{School of Physics and Astronomy, University of Minnesota, Minneapolis, MN 55455, USA}
\affiliation{ISC-CNR and Department of Physics, Sapienza University of Rome, Piazzale Aldo Moro 2, I-00185, Rome, Italy}
\author{Jonathan Ruhman }
\affiliation{Department of Physics, Bar Ilan University, Ramat Gan 5290002, Israel}
\author{Rafael M. Fernandes}
\affiliation{School of Physics and Astronomy, University of Minnesota, Minneapolis, MN 55455, USA}
\begin{abstract}

Doped SrTiO$_3$, one of the most dilute bulk systems to display superconductivity, is perhaps the first example of an unconventional superconductor, as it does not fit into the standard BCS paradigm. More than five decades of research has revealed a rich temperature-carrier concentration phase diagram that showcases a superconducting dome, proximity to a putative quantum critical point, Lifshitz transitions, a multi-gap pairing state and unusual normal-state transport properties. Research has also extended beyond bulk SrTiO$_3$, ushering the new field of SrTiO$_3$-based heterostructures. Because many of these themes are also featured in other quantum materials of contemporary interest, recent years have seen renewed interest in SrTiO$_3$. Here, we review the challenges and recent progress in elucidating the superconducting state of this model system. At the same time that its extreme dilution requires to revisit several of the approximations that constitute the successful Migdal-Eliashberg description of electron-phonon superconductivity, including the suppression of the Coulomb repulsion via the Tolmachev-Anderson-Morel mechanism, it opens interesting routes for alternative pairing mechanisms whose applicability remains under debate. For instance, pairing mechanisms involving longitudinal optical phonons have to overcome the hurdles created by the anti-adiabatic nature of the pairing interaction, whereas mechanisms that rely on the soft transverse optical phonons associated with incipient ferroelectricity face challenges related to the nature of the electron-phonon coupling. Proposals in which pairing is mediated by plasmons or promoted locally by defects are also discussed. We finish by surveying the existing evidence for multi-band superconductivity and outlining promising directions that can potentially shed new light on the rich problem of superconductivity in SrTiO$_3$.

\end{abstract}
\maketitle

\tableofcontents

\section{Introduction}\label{Sec:intro}
Doped SrTiO$_3$ (STO) was discovered to superconduct in 1964~\cite{Schooley1964}, only 7 years after the BCS theory and 4 years after the Eliashberg theory~\cite{Eliashberg1960} were published. This was the first example of a superconductor that cannot be described by the BCS-Eliashberg paradigm, and may therefore be identified as the first unconventional superconductor. Its unconventional character does not refer to a non-trivial gap symmetry, but to the fact that it does not seem to arise from the most standard electron-phonon pairing mechanism. Motivated by the idea that semiconductors could be useful systems for studying superconductivity, the theoretical work by Cohen~\cite{Cohen1964RevModPhys,Cohen1964superconductivity} prompted the discovery of superconductivity in STO and other degenerate semiconductors~\cite{bustarret2008superconducting,Hein1964}.
Yet, despite more than 50 years of intense experimental and theoretical activity, the origin of superconductivity in this material remains an open problem in quantum condensed matter physics. 

Recently, several groups have revisited this fascinating problem, applying advanced experimental and theoretical techniques that were developed over the past decades to study quantum materials. On the theoretical front, one of the main challenges is on how to extend the very successful Migdal-Eliashberg theory of electron-phonon superconductors \cite{Marsiglio2019} to such a dilute system. Establishing such a theoretical framework would clearly have an impact on the understanding of dilute superconductivity that emerges from other lightly-doped semiconductors and semimetals, such as Bi \cite{Prakash2017}, YPtBi \cite{Butch2011}, and PbTe \cite{Matsushita2006}. In this regard, we note that the generalization of the Migdal-Eliashberg formalism to unconventional superconductors has been a very active area of research, although the focus has been mostly on pairing mechanisms that do not involve electron-phonon coupling (see, e.g. \cite{Abanov2003}).

\begin{figure}
 \begin{center}
    \includegraphics[width=0.75\linewidth]{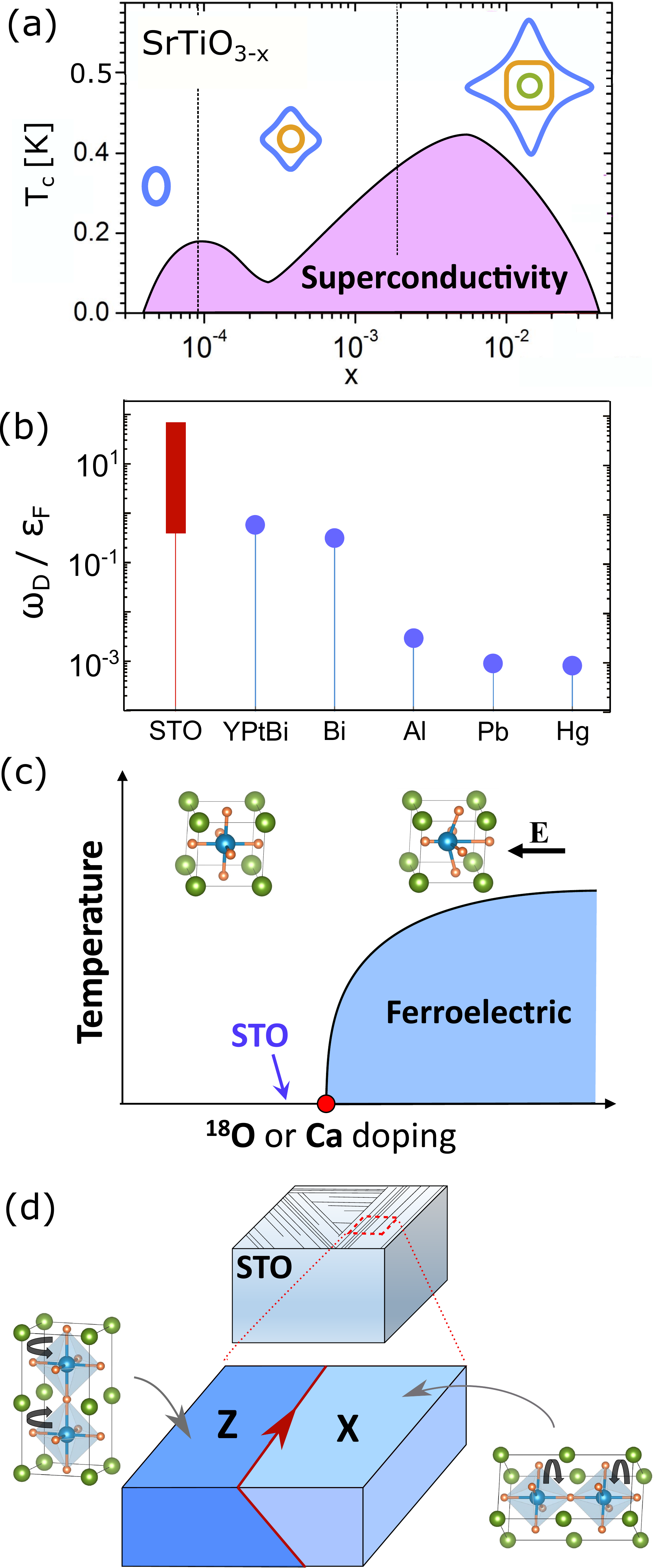}
 \end{center}
\caption{(a) $T_c$ as a function of density in units of electrons per formula unit (schematic, based on Ref. \cite{Lin20133}). Dashed vertical lines mark Lifshitz transition points. Insets are the Fermi surfaces around the $\G$-point. (b) The Migdal ratio $\w_D / \e_F$ for different materials. The values for STO, YPtBi and Bi are taken from Refs.~\cite{duran2008specific,Meinert2016Unconventional,DeSorbo1958,Tediosi2007Charge}. The red strip corresponds to $\e_F$ in the density range of the SC dome~\cite{Lin20133}. (c) STO naturally lies on the verge of a ferroelectric quantum critical point (QCP) separating a paraelectric and a ferroelectric phase. The transition is structural, between tetragonal and non-centrosymmetric structures (see inset). Pristine STO is paraelectric. Doping with Ca or $^{18}$O drives STO through the transition. (d) The cubic-to-tetragonal structural antiferrodistortive transition at 105K. The order parameter spontaneously breaks the cubic symmetry, selecting an axis among X, Y and Z. As a result, at low temperatures, STO is heterogeneous and filled with X, Y and Z domains. Conductivity is enhanced at the domain walls separating these regions (indicated by the red arrow). Whether these domains play an important role in superconductivity remains an open question.  }
 \label{fig:intro}
\end{figure}

One of the appeals of studying STO as a model system is that its phase diagram shares important features with a variety of quantum materials that are at the forefront of research in superconductivity~\cite{bednorz1988perovskite}. This includes the emergence of a superconducting dome as a function of carrier concentration, the existence of quantum fluctuations, the interplay with structural instabilities, and the correlation with unusual normal-state transport properties. In this paper, we provide an overview of what we consider to be the most fascinating challenges for the elucidation of superconductivity in STO, discussing possible paths forward. Our focus will thus be in the superconducting state of STO, and on why it still defies our understanding. Of course, given the many decades of research dedicated to this problem, important topics will not be covered here. We refer the interested reader to other reviews on STO (see e.g. Collignon \etal \cite{Collignon2019metal}).

\subsection{Brief Summary of the Essential Experimental Results}
\begin{figure}
\begin{center}
   \includegraphics[width=0.8\linewidth]{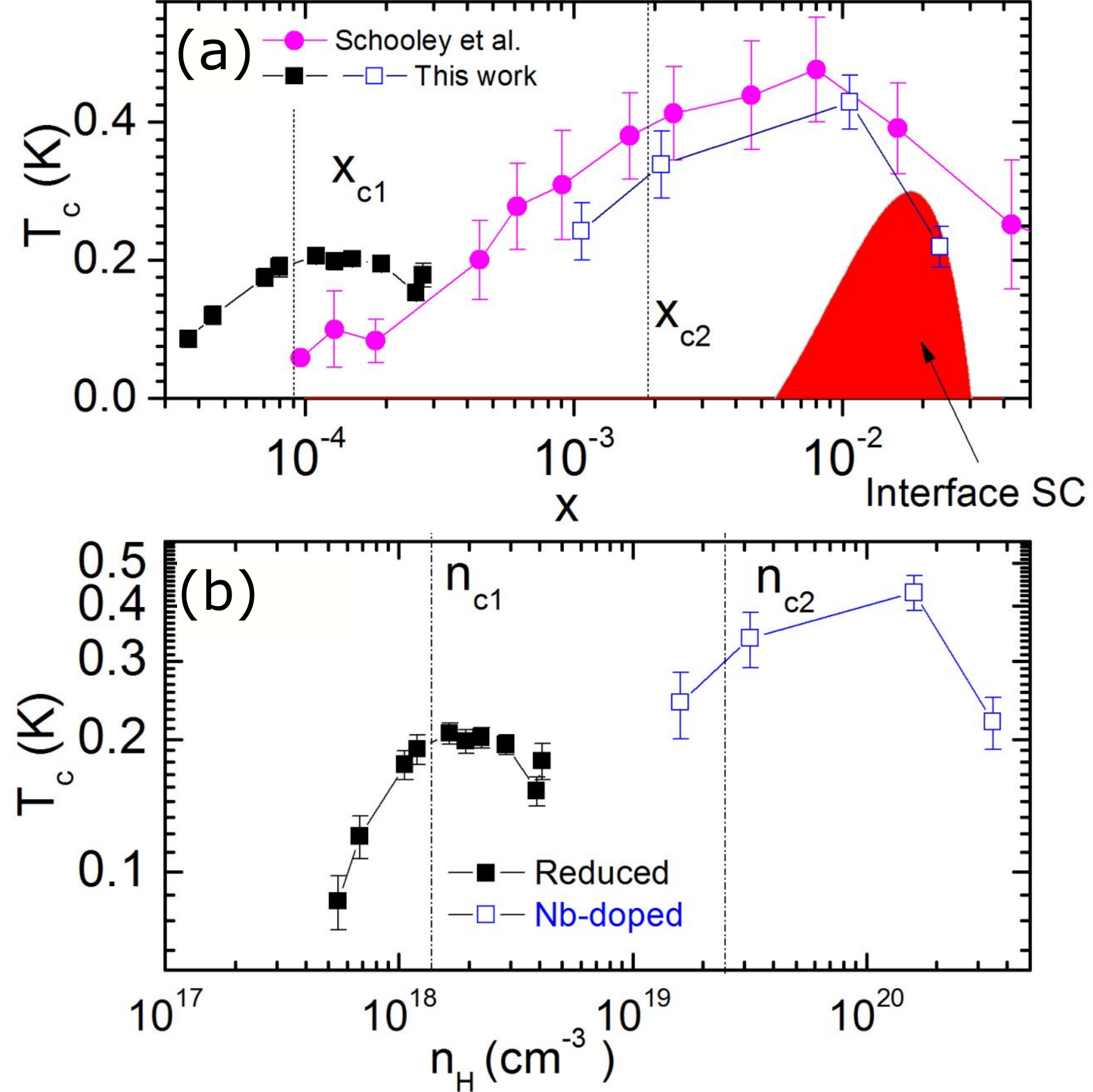}
 \end{center}
\caption{The superconducting phase diagram of doped STO, from Ref. \cite{Lin20133}. The lower concentrations are achieved via O reduction and the higher concentrations, via Nb doping. The dashed lines mark the Lifshitz transitions where additional bands cross the Fermi level. The red superconducting dome refers to the LaAlO$_3$/SrTiO$_3$ interfaces. The original data by Schooley \etal is shown by the pink circles \cite{Schooley1964}. Figure reproduced with permission from Ref. \cite{Lin20133}. Copyright 2014 by the American Physical Society.}
 \label{fig:Behnia}
\end{figure}

We start with a brief summary of the milestones in the experimental literature that provide the essential information required to set up the problem of superconductivity in STO. Following the experimental discovery by Schooly \etal \cite{Schooley1964}, Koonce \etal published an extended data set in 1967  \cite{Koonce1967} showing that the superconducting $T_c$ 
exhibits a dome as a function of density, starting around $n \sim 10^{18}$ cm$^{-3}$ and ending above $n \sim 10^{21}$ cm$^{-3}$. This corresponds to a tiny fraction $x$ of electrons per unit cell, as shown schematically in Fig. \ref{fig:intro}(a) (for the actual experimental phase diagram, see Fig. \ref{fig:Behnia}).  Strikingly, $T_c$ depends very weakly on the density in this large window and stands roughly at a few hundreds mili-Kelvin.

Since undoped STO is a semiconductor, it is useful to contrast these numbers with the theory for superconductivity of doped semiconductors developed by Gurevich, Larkin, and Firsov (GLF theory) \cite{Gurevich1962}. In this theory, the role of the Debye frequency $\w_D$ in the usual BCS theory of superconductivity is replaced by the longitudinal optical phonon frequency, $\w_L$. In STO, $\w_L$ is of the order of 100 meV~\cite{Kamaras1995,Zhou2018}.
In contrast, the Fermi energy varies between 2 to 60 meV in the density range specified above~\cite{Lin20133}, which clearly violates the conditions of applicability of the standard Migdal-Eliashberg theory ($\w_D \ll \e_F$) \cite{Marsiglio2019}. Actually, STO has the highest $\w_D/\e_F$ ratio among all superconductors, including other dilute systems such as Bi [see Fig. \ref{fig:intro}(b)].

Recently, the interest in the superconducting state of STO has re-emerged and a number of key results have been obtained. First, the low-density bound on superconductivity was pushed to even lower densities $\sim  10^{17}$ cm$^{-3}$~\cite{Lin20133,Bretz2019}, where quantum oscillations indicate that the Fermi surface is still sharp~\cite{Lin2014}. They also indicate multiple Lifshitz transitions, i.e. transitions in which the number of bands crossing the Fermi level increases [see insets of Fig. \ref{fig:intro}(a)]. The superconducting state was shown to be robust against disorder, which was interpreted as a signature of $s$-wave pairing~\cite{Lin_irraditiona_2015}. Tunneling experiments at high densities showed that the ratio of $k_B T_c/\Delta$ fits the prediction of weak-coupling BCS theory, although the coupling to longitudinal optical modes (the modes considered in GLF theory) is very strong~\cite{Swartz2018}. Finally, the bulk superconducting transition temperature, as identified from magnetic susceptibility and specific heat measurements, was found to deviate significantly from the resistive superconducting transition temperature, which is an indication of filamentary superconductivity~\cite{Collignon2019metal,Collignon2017}.

In addition to superconductivity, STO also has a huge static dielectric constant $\ve_0 \approx 2\times 10^4$~\cite{weaver1959}. The large $\ve_0$ is a manifestation of a nearby paraelectric-ferroelectric quantum phase transition ~\cite{Muller1979,Rowley2014}. Classically, the lowest energy configuration of the crystal, as determined by first-principles, involves a ferroelectric distortion of the oxygen octahedron [see insets of Fig.~\ref{fig:intro}(c)]. However, long-range order is prevented by quantum fluctuations. Consequently, there is a low-energy 1 meV optical phonon mode~\cite{Shirane1969LAttice,Vogt1981,Kamaras1995,Vogt1995}, which remains soft even when the system is doped with carriers~\cite{bauerle1980soft,vanderMarel2008,Crandles1999}. 

The existence of a soft bosonic mode, and of a putative ferroelectric quantum critical point (QCP) associated to it, has recently motivated many theoretical and experimental studies \cite{Chandra2017}. Edge \etal proposed that the pairing interaction in dilute STO is mediated by critical ferroelectric fluctuations~\cite{Edge2015}. This idea has similarities to proposals that magnetic quantum critical fluctuations provide the pairing glue in strongly correlated materials ~\cite{Taillefer2010scattering,Scalapino2012common}. As a result, the interplay between the ferroelectric transition and superconductivity has been intensively studied experimentally by tuning superconducting STO through the ferroelectric critical point. This has been accomplished by Ca doping~\cite{rischau2017ferroelectric}, oxygen isotope substitution~\cite{stucky2016isotope}, La doping~\cite{tomioka2019enhanced}, hydrostatic pressure~\cite{rowley2018superconductivity} and strain~\cite{herrera2018strainengineered,Harter2019} [see Fig.~\ref{fig:intro}(c)].
In all cases, an enhancement of $T_c$ when approaching the critical point has been observed. However, more results are needed to establish whether a superconducting dome is formed around the QCP, as theory predicts.  

The softness of the oxygen sublattice in the STO crystal is also manifested in a structural antiferrodistortive (AFD) transition from cubic to tetragonal as the temperature is lowered below $T_{\mathrm{AFD}} = 105$ K. The existence of this transition causes another optical phonon branch to become soft near the zone boundary~\cite{Shirane1969LAttice}. 
More importantly, the spontaneous symmetry breaking below the transition leads to a noticeable heterogeneous structure [see Fig.~\ref{fig:intro}(d)]. At low temperatures, different tetragonal domains proliferate in the sample~\cite{dec2004electric}, separated by domain walls~\cite{honig2013local}. Kalisky \etal found that the conductivity is enhanced along these defects~\cite{kalisky2013locally}, which have also been associated to the observation of a locally higher $T_c$~\cite{Pai2018One,noad2019modulation}.

In 2004, Othomo and Hwang~\cite{ohtomo2004high} discovered that a two-dimensional electron gas forms at the interface between a thin LaAlO$_3$ film  and a pristine SrTiO$_3$ substrate. The electronic properties of the two-dimensional gas resemble those of bulk STO. Importantly, the 2D state is also superconducting, with a $T_c$ value similar to that of the bulk \cite{Reyren2007}. A superconducting dome also emerges as a function of density~\cite{caviglia2008electric}, which can be controlled continuously via gating (see Fig. \ref{fig:Behnia}). The similarity between the transition temperatures of the bulk and of the interface raises the question of whether the two systems share the same microscopic pairing mechanism. 

The normal state of STO is also unusual, as its resistivity displays an unexpected temperature dependence. At low temperatures, it exhibits a strong $T^2$ behavior~\cite{vanderMarel2011,lin2015scalable,Stemmer2018}, which is not expected due to the tiny Fermi surface~\cite{lin2015scalable}. The $T^2$ coefficient was also argued to violate the Kadawoki-Woods scaling~\cite{McCalla2019,Hussey2005}. At higher temperatures, the $T^2$ behavior switches to a $T^3$ behavior, and the resistivity becomes so large that it cannot be explained by standard Drude theory~\cite{lin2017metallicity}. An interesting unresolved question is whether these unconventional transport properties might be related to the soft modes that have been proposed to provide the pairing interaction at low temperatures.

\subsection{Challenges for a Microscopic Description of Low-Density Superconductivity}\label{Sec:Issues}
As shown in Fig. \ref{fig:Behnia}, STO exhibits superconductivity at very low density $n\sim 10^{17}$ cm$^{-3}$\cite{Schooley1964,Lin2014,Lin20133,Bretz2019}, corresponding to a very small Fermi energy ($\sim 1$ meV). In conventional Migdal-Eliashberg theory, the electronic attraction comes from the exchange of longitudinal phonons, whose typical frequencies are much smaller than the Fermi energy~\cite{Margine2013}. The electrons and phonons couple to each other through the retarded, short-ranged deformation potential generated by the displacement of the atoms in the crystal~\cite{hamaguchi2010basic}.
Extensions of the standard Migdal-Eliashberg theory to explain the superconductivity in STO requires addressing several issues, such as:

\noindent
(i) \textbf{The Migdal criterion is violated --} In contrast to conventional superconductors, where the ratio between the Fermi energy and the phonon frequency  is of the order of $\w_D / \e_F \sim 10^{-2}\,-\,10^{-3}$, the ratio in STO is $\w_D / \e_F\sim 1\,-\,10^2$ [see Fig.~\ref{fig:intro}(b)]. The system is said to be in the anti-adiabatic regime, outside the range where the Migdal approximation can justify the omission of vertex and other corrections \cite{migdal1958interaction,Grabowski1984,Sadovskii2019_1}. As a result, which Feynman diagrams must be included in a generalized Migdal-Eliashberg theory remains an open question. 

\noindent
(ii)  \textbf{The Coulomb repulsion may not be efficiently suppressed --} 
 In standard Migdal-Eliashberg theory, the high-energy cutoff is usually set at the Fermi energy, below which the Coulomb repulsion is short-ranged and essentially independent of frequency (since the plasmon frequency is much larger than $\e_F$). The use of a sharp cutoff is justified by the fact that the interaction is nearly frequency independent over the wide range of frequencies $\w_D < \w < \e_F$, which makes the end result depend only logarithmically on the cutoff (the Tolmachev logarithm, see ~\cite{Tolmachev1958}). In this case, the impact of the Coulomb repulsion on $T_c$ is strongly suppressed by pair excitations in the range $\w_D < \w < \e_F$, giving rise to the so-called Anderson-Morel Coulomb pseudo-potential \cite{Morel1962}. In the low-density regime, however, the effectiveness of this Tolmachev-Anderson-Morel mechanism in suppressing the Coulomb repulsion is much less obvious \cite{Sadovskii2019_2}, as the pairing interaction is expected to display a significant frequency-dependence for energies near $\e_F$. This also causes the value of $T_c$ to depend strongly on the energy cutoff, making its precise location important \cite{ruhman2019comment,wolfle2019reply}.  

\noindent
(iii) \textbf{The density of states is very small --} In three-dimensional systems such as STO, the density of states $\nu$ vanishes in the limit of zero density. As a result, the BCS coupling strength $\l = \nu V_0$ arising from the phonon-mediated interaction $V_0$ is suppressed by the same amount as the density of states. Given that this coupling goes in the exponent, it makes the predicted
$
T_c \sim \w_D\, \exp \left[-{1/ \nu V_0}  \right]
$
immeasurably small. Of course, in the very dilute regime, the BCS logarithm from which the equation above is derived disappears, and the attractive interaction must overcome a threshold value to cause pairing. As a result, the standard Migdal-Eliashberg approximation of considering only states near the Fermi level needs to be revisited, as the gap function may depend substantially not only on frequency, but also on momentum \cite{Gastiasoro2019}. 

\subsection{Possible Mechanisms for Superconductivity}
In the previous subsection we have argued that the standard Migdal-Eliashberg approach cannot be applied in a straightforward way to describe the superconducting state of STO. In this subsection, we discuss several ideas, some of which attempt to generalize the Eliashberg theory, that have been put forward to circumvent the issues (i)-(iii) discussed above. 

\subsubsection{Long-range electron-phonon interaction}  \label{sec:dynamical_coulomb}
We start with issue number (iii), and with the seminal work of Gurevich, Larkin and Firsov (GLF)~\cite{Gurevich1962}. In this paper the authors studied the bounds on superconductivity in semiconductors. The main premise was to point out that long-range attractive interactions can cause a relatively high transition temperature in spite of the low density of states. For instance, let us consider as a toy model the case of an attractive Coulomb interaction $V\simeq -4\pi e^2/\ve q^2$. In such a case, the dimensionless coupling strength characterizing this interaction is the dimensionless density  $r_s = {\a    /  a_B k_F } $, where $a_B$ is the Bohr radius and $\a = \left({9\pi  / 4} \right)^{1\over 3}$. Thus, the coupling strength is enhanced, rather than suppressed, in the low-density limit. 

GLF pointed out that such an attractive interaction appears naturally in polar crystals through the exchange of longitudinal optical (LO) phonons (see Section \ref{Sec:Pairing} for an extended discussion). 
The exchange of these phonons renormalizes the Coulomb repulsion and adds a dynamical contribution from the lattice in the long-wavelength limit. For example, in the case of a single LO mode we obtain the screened Coulomb interaction
\begin{align}\label{eq:screened_V_Coulomb}
V_C (\w,\bs q) &= {4 \pi e^2 \over\ve_c (\w,\bs q\rightarrow 0) q^2}  \\
&={4\pi e^2 \over \ve_\infty q^2}\left[ 1 - \left( {1\over \ve_{\infty}} - {1\over \ve_0} \right) {\w_L^2\over \w_L^2 -\w^2} \right]\nn
\end{align}
where $\ve_0$ and $\ve_\infty$ are the low and high frequency dielectric constants and $\w_L$ is the LO phonon frequency. Given that $\ve_0 > \ve_\infty$ the second term is attractive. The long-range character of this term, manifested by its $1/q^2$ dependence, arises from the absence of electronic screening in the very dilute regime.

 GLF argued that as long as $\w_L$ is much smaller than the Fermi energy $\e_F$, the Migdal criterion is obeyed and the standard BCS approach can be applied. They obtained a non-negligible $T_c$ in spite of the small density of states typical of a doped semiconductor. In this analysis the Coulomb repulsion Eq.~\eqref{eq:screened_V_Coulomb} is transmuted into attraction via the standard Coulomb pseudo-potential method ~\cite{Morel1962}. Thus, while this yields a possible solution of issue number (iii), it does not address issues (i) and (ii). 

More generally, we may consider the entire dynamics of the dielectric constant, including contributions from the electronic liquid itself (i.e. the plasmonic modes) in addition to the LO phonons.  This was first studied by Takada in 1978~\cite{takada1978plasmon}. 
The straightforward generalization to include plasmons raises a few theoretical issues. The main one is that the plasmon resonance typically occurs at an energy scale comparable to the Fermi energy and, therefore, suffers from issues (i) and (ii) above ~\cite{Grabowski1984,ruhman2016superconductivity}. Despite these drawbacks, Takada solved the Eliashberg equations \cite{takada1978plasmon,takada1980theory,takada1992plasmon,takada1993s} and found a solution in the large $r_s$ limit, which remains an important observation.

Thus, the dynamically screened Coulomb repulsion provides an effective pairing mechanism in systems with low density of states. It is especially relevant to polar crystals where in addition to the plasmonic mode there are also the LO modes considered by GLF. Thus, this mechanism seems highly relevant for STO. 
However, for the theory to be controllable, a pseudo-potential mechanism must be invoked to avoid the strong Coulomb repulsion. The usual pseudo-potential mechanism requires that the polar mode frequencies (LO phonons and plasmons) are smaller than the Fermi scale, which  does  not apply to STO across the entire range of concentrations where superconductivity is seen. We will return to this mechanism in Sec. \ref{Sec:Pairing}. 

\subsubsection{Soft bosonic modes}\label{sec:soft_modes_intro}
The most important conclusion from GLF theory is that long-range attractive interactions, intrinsic to low-density systems, provide a possible pairing glue. As we saw, however, the dynamically screened Coulomb repulsion proposed in GLF theory becomes problematic if the density is too low, because the Fermi energy always becomes smaller than the LO phonon frequency. Thus, it is important to identify alternative sources for long-range attractive interactions.  

On quite general grounds, long-range interactions arise when soft bosonic modes couple to the states at the Fermi surface with zero momentum transfer. The interaction mediated by such a mode is given by 
\be \label{eq:critical_mode_int}
V_S^{\a\b\g\d}(\bs k,\bs k';\w,\bs q) = -{A_{\a\b}^i(\bs k;\bs q) A_{\g\d}^j(\bs k';-\bs q) \chi_{ij}(\w,\bs q)}
\ee
where $A_{\a\b}(\bs k ;\bs q)$ is the coupling matrix element of electronic states at $|\bs k,\a\rangle$ and $|\bs k+\bs q,\b\rangle$, while $\chi_{ij}(\w,\bs q)$ is the bosonic propagator. Here, Greek letters ($\alpha$, etc) denote spin and Latin letters ($i$, etc) denote the components of the bosonic field. When the bosonic mode is soft, the static susceptibility $\chi$ diverges, usually according to $1/q^2$ in the zero-frequency limit. Thus, as long as $A$ remains finite at vanishing momentum transfer, the resulting interaction \eqref{eq:critical_mode_int} is long-range. In some cases, the attraction may even be in a non $s$-wave channel. 

These conditions are naturally fulfilled in two well known cases: (i) at quantum critical points; and (ii) inside an ordered phase where a continuous symmetry has been broken and a Goldstone mode exists. It is important to note that for the second case a sufficient condition for $A$ to remain finite at $q\to 0$ is that the generators of the symmetry that is being broken do not commute with the momentum operator~\cite{Watanabe16314}.

Acoustic phonons are essentially Goldstone modes, and therefore can potentially lead to long-range interactions of the form \eqref{eq:critical_mode_int}. However, because they result from the breaking of translational symmetry, which is generated by the momentum itself, they do not couple to the electronic density at $q\to 0$. Indeed, the electronic coupling to acoustic phonons is given by the gradient, i.e. $A(q) = {-i q D/\sqrt{\rho }}$ at small $q$, where $D$ is the deformation potential and $\rho$ is the mass density.   

A natural candidate for a soft mode in STO, as explained above, is the transverse optical phonon mode associated with quantum ferroelectric fluctuations. 
Edge \etal \cite{Edge2015} have proposed that the bosonic fluctuations close to the ferroelectric quantum critical point are responsible for the superconducting dome in lightly doped STO. The proposal has spurred much interest~\cite{kedem2016unusual,wolfle2018superconductivty,Arce-Gamboa2018quantum,rowley2018superconductivity,Kanasugi2018spin,kedem2018novel,ruhman2019comment,wolfle2019reply,van2019possible} and experimental activity~\cite{stucky2016isotope,rischau2017ferroelectric,tomioka2019enhanced,rowley2018superconductivity,van2019possible,herrera2018strainengineered}. We will discuss more about these ideas in Sec. \ref{Sec:Pairing}.   

\subsubsection{The rise and fall of intervalley phonons}\label{sec:intervalley-intro}
In 1964, Cohen identified an elegant way for electrons to couple strongly to phonons despite the gradient coupling discussed in Sec. \ref{sec:soft_modes_intro}~\cite{Cohen1964superconductivity,Koonce1967}. In particular, he pointed out that when there are multiple small Fermi pockets (or valleys) separated by momenta comparable to the Brillouin zone (BZ) size, then phonon processes involving pair scattering between the pockets can carry a large momentum transfer. This can lead to a BCS interaction of the form 
\be V_S(\w,\bs k - \bs p) c_{\bs k, 1}^{\dag} c_{-\bs k, 1 }^{\dag} c_{-\bs p, 2} c_{\bs p, 2} \ee
where $c_1$ and $c_2$ denote the electronic states on different pockets. Since the momentum transfer $\bs q=\bs k- \bs p$ is comparable to the BZ size, even the gradient coupling can be large. 
Such a soft phonon mode exists in STO due to the antiferrodistortive transition at 105 K~\cite{Fleury1968}.

However, it was later understood that the electronic states in lightly doped STO lie in a single or in multiple pockets all centered around the $\G$-point of the BZ ~\cite{Mattheiss1972,ALLEN1973411}. This rules out the option of intervalley processes as promoting pairing.
Later on, Ngai~\cite{Ngai} proposed that the gradient coupling can be avoided also if two-phonon processes are considered. However, this is a higher order term, which is typically small.

\subsubsection{Other possible mechanisms}
So far, we have focused on possible pairing mechanisms based on long-range attraction. 
In this subsection, we will briefly discuss some alternative ideas to bypass some of the issues (i), (ii) and (iii), which have not been fully explored yet. 

One elegant manner in which the issue of low density of states (iii) can be circumvented is by reducing the dimensionality. This is because in one and two dimensions the density of states is not necessarily reduced in the low-density limit. It is interesting to point out that Kalisky \etal \cite{kalisky2013locally} found enhanced conductivity on the domain walls between different tetragonal distortion orientations [see Fig.~\ref{fig:intro}(d)]. These were later speculated to be the source of lower-dimensional superconductivity in STO by Pai \etal \cite{Pai2018One}. It is not clear if the rich phenomenology of superconductivity in STO can be explained by two-dimensional superconductivity residing on domain walls, but it is definitely a promising direction of research. 

Another line of reasoning is that of localized modes. Gor'kov has conjectured that the attractive interactions in STO are instantaneous \cite{gor2015phonon}, i.e. the Coulomb repulsion is over-screened, due to the multiple longitudinal modes present in STO. Such an interaction would avoid the issue of introducing a Coulomb pseudo-potential to reduce the Coulomb suppression of $T_c$ \cite{Tolmachev1958,Morel1962} [issue (ii)]. Under this assumption, the phase diagram of doped STO was reproduced in Ref. \cite{gor2015phonon}.
However, such an over-screening of the Coulomb interaction is not usually possible in a classical dielectric medium~\cite{ruhman2016superconductivity}. In a follow-up paper~\cite{Gorkov2017}, Gor'kov proposed that if the contribution from local polar defects is added, in addition to the screening from the perfect lattice, such an instantaneous attraction can be generated. While the suitability of this idea to STO remains to be established, it is interesting to connect it to a recent experiment where the existence of these defects and their strong interaction with the electrons was demonstrated~\cite{wang2019charge}. 

A third interesting proposal for instantaneous attraction, that is somewhat related to the previous one, has been raised by Geballe \cite{Matsushita2006,kivelson2019physics}. The idea is that certain dopants that have ``skipping valence" naturally produce strong local attractive interactions on the dopant sites. Such a situation may occur in oxygen vacancies that skip directly between Ti$^{3+}-$V$_\text{O}-$Ti$^{3+}$ and standard Ti$^{4+}-$V$_\text{O}-$Ti$^{4+}$. In this situation, the attraction is generated locally and its strength would thus depend on the fraction of oxygen vacancies present in the lattice.

\section{Normal state properties}\label{Sec:NormalState}

\subsection{Lattice Properties}

\begin{figure}
 \begin{center}
    \includegraphics[width=0.9\linewidth]{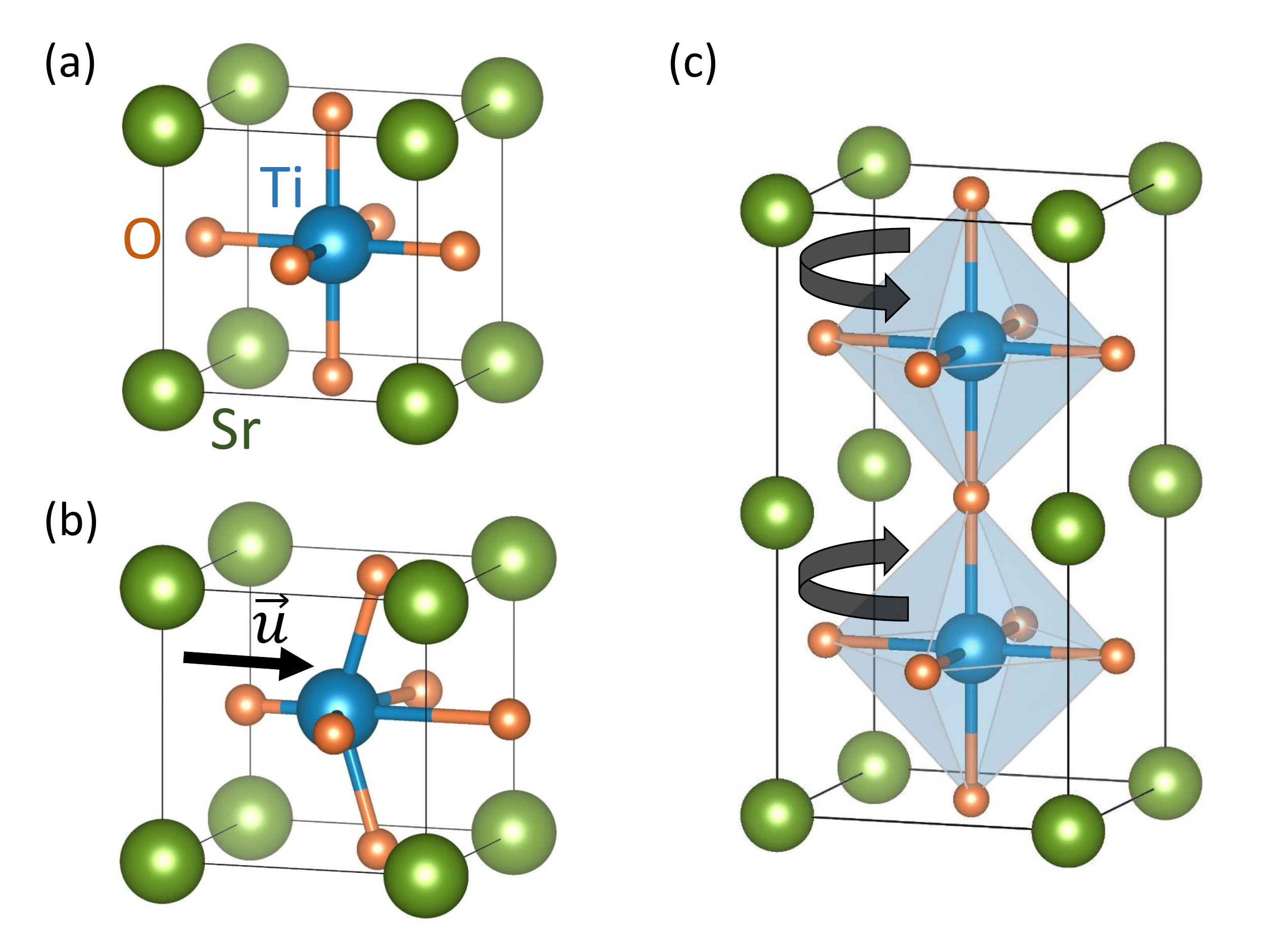}
    \includegraphics[width=\linewidth]{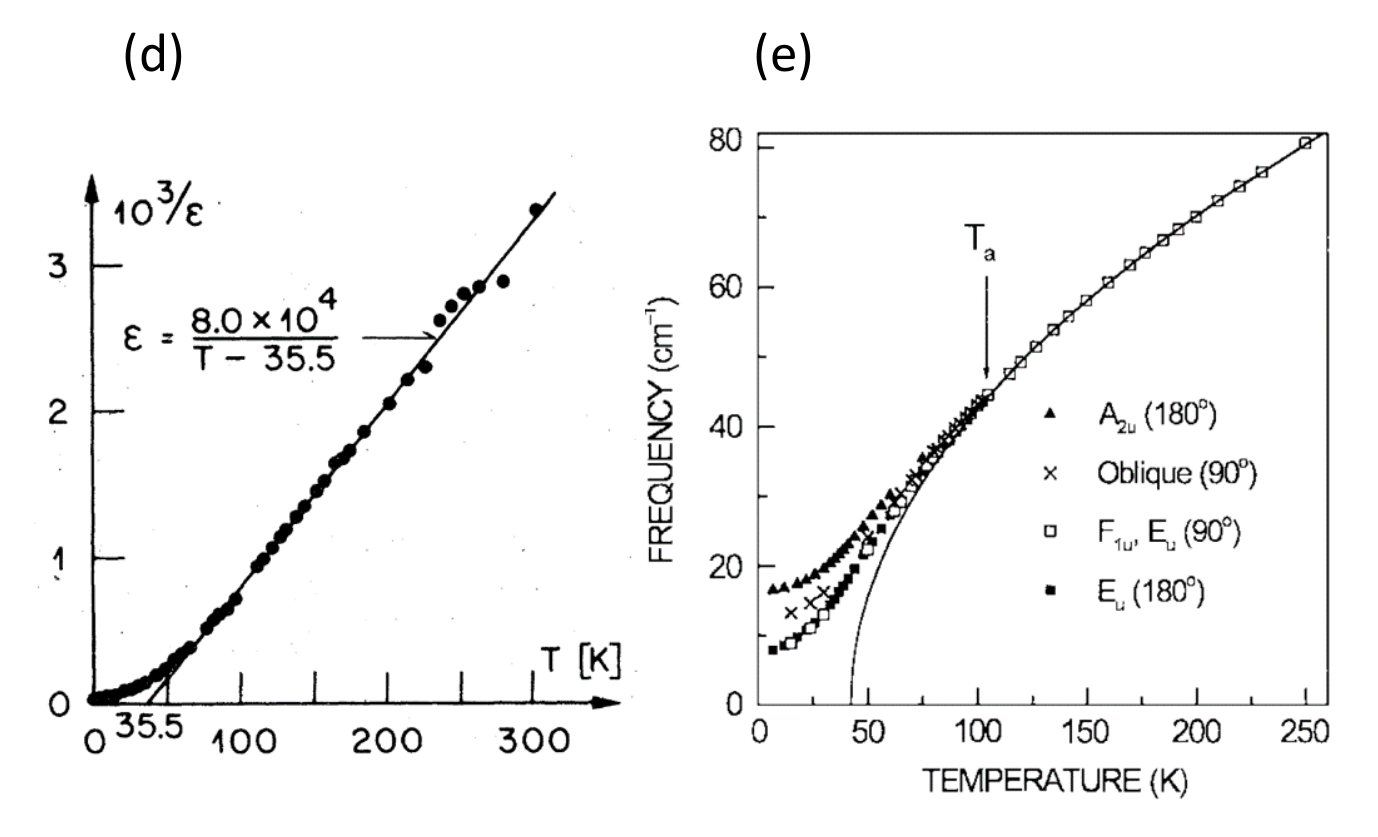}
 \end{center}
\caption{Lattice properties of STO. Panel (a) represents the unit cell of the cubic paraelectric phase. Panel (b) illustrates the lattice distortion associated with ferroelectricity. Panel (c) shows the unit-cell doubling taking place at the antiferrodistortive  transition (cubic-to-tetragonal) due to staggered rotation of the oxygen octahedra. (d) Inverse dielectric constant $10^3/\varepsilon(T)$ as function of temperature. From Ref.~\cite{Muller1979}. (e) Softening of the TO mode $\omega_{T,1}(T)$ associated with ferroelectricity extracted from hyper-Raman measurements~\cite{Yamanaka2000}. The splitting of the mode at $T_{\mathrm{AFD}}=105$ K is due to the antiferrodistortive transition. Panel (d) reproduced with permission from Ref. \cite{Muller1979}. Copyright 1979 by the American Physical Society. Panel (e) reproduced with permission from Ref. \cite{Yamanaka2000}. Copyright 2000 by EDP Sciences.}\label{fig:FE_diag}
\end{figure}

STO has a cubic perovskite structure ($Pm\bar{3}m$) at room temperature, as shown in Fig.~\ref{fig:FE_diag}(a). Five atoms per unit cell (Sr$^{2+}$, Ti$^{4+}$ and 3 O$^{2-}$) give rise to 3 acoustic-phonon branches and 12 optical branches at a general $\bs k$ point in the Brillouin zone. At the zone center ($\bs q=0$), a group theory analysis finds three \textit{polar} optical $\Gamma^-_{15}$ modes in addition to the acoustic modes at $\w=0$ and a triply degenerate optical nonpolar/normal $\Gamma^-_{25}$ mode (see for example Ref.~\cite{Cowley1964}). 

The ionic displacement associated with the polar modes produces an electric dipole moment, as illustrated in Fig.~\ref{fig:FE_diag}(b), due to the relative motion of the cation (either Ti or Sr) with respect to the anion (oxygen).
Due to this coupling between the lattice displacement and the electric polarization, the polar phonon modes are subject to long-range Coulomb interactions. As a result, each of the polar modes splits, at the zone center, into one longitudinal optical (LO) mode with frequency $\w_{L,j}$ and one doubly degenerate transverse optical (TO) mode with frequency $\w_{T,j}$. The frequencies of these three pairs of LO and TO phonons have been extensively studied in the literature~\cite{Vogt1981,Kamaras1995} and are summarized in Table~\ref{tab:numbers}. 
The large TO/LO splittings indicate the strong polar character of this material.

\begin{table}
  \begin{center}
    \caption{(a) Frequencies of the polar optical phonons $\Gamma^-_{15}$ at room temperature extracted from infrared~\cite{Kamaras1995} and Hyper-Raman~\cite{Vogt1981} experiments. The numbering of the modes is in order of increasing frequencies. The frequency values for the soft TO1 phonon appearing in brackets correspond to the frequencies at $T=4$K and $T=300$K, respectively. (b) Parameters (in meV) of the tight-binding model of Eq.~\eqref{eq:TB-model}, fitted to the DFT electronic structure. The mass enhancement of $2$ is doping independent~\cite{McCalla2019}, see also Fig.~\ref{fig:sommerfeld}.}
    \label{tab:numbers}\vspace{0.2cm}
    \begin{tabular}{|c|c|c|c|} 
    \hline
    \multicolumn{4}{|c|}{(a) Lattice properties} \\
    \hline\hline
     TO mode & $\w_i$ [meV] & LO mode & $\w_i$ [meV]\\
      \hline
   	 TO1 &  $(1,11.3)$ & LO1 & $21.3$ \\
      \hline
      TO2 &  $21.7$ & LO2 & $58.7$\\
      \hline
      TO3 &  $67.4$ & LO3 & $98.1$ \\
      \hline
      \hline\hline
    \multicolumn{4}{|c|}{(b) Electronic properties} \\
    \hline\hline
    $t_1$  & $t_2$  &  $\xi$ &  $\Delta$ \\
      \hline
     615 & 35 & 19.3 & -2.2 \\
      \hline
      $\epsilon_0$ & $\mu(n_{c1})$  & $\mu(n_{c2})$ & $m^*/m^*_{\mathrm{th}}$ \\
      \hline
      12.2 & 4.7 & 31.8 & 2 \\
      \hline
    \end{tabular}
  \end{center}
\end{table}

The dielectric function of STO can be approximated using a generalized Lyddane-Sachs-Teller (LST) relation~\cite{LST1941,cochran1962}  
\be\label{eq:LST}
{\ve_p}(\w,q)=\ve_\infty\prod_{j=1}^3\frac{\w^2_{L,j}-\w^2}{\w^2_{T,j}-\w^2}\,.
\ee 
where the optical frequencies $\w_{T,j}$ and $\w_{L,j}$ are listed in Table \ref{tab:numbers}. They are approximately constant, except for the soft mode $\w_{T1}$, which is very sensitive to perturbations such as temperature $T$, doping $n$, and external electric fields $E$. We can summarize these dependencies (at low temperatures) in the phenomenological equation
\be\label{eq:wT}
\w_{T,1}^2(q,T,E,n) = \w_0^2+(c_T q)^2+(\g_T T)^2+(\g_E E)^2+\g_n n
\ee
where $c_T \approx 5\,\mrm{meV}\,\mrm{nm}$~\cite{Shirane1969LAttice}, $\g_T \approx 6.3\times10^{-2}\,\mrm{meV/K}$~\cite{Rowley2014}, $\g_E \approx 10^{-3}\,\mrm{meV\,cm/V}$~\cite{worlock1967electric}~ and $\g_n \approx 1.16 \times 10^{-17}\,\mrm{cm}$~\cite{bauerle1980soft,Crandles1999}. Here, $q$ is the momentum. To obtain the corresponding dependence of the dielectric constant, one simply substitutes Eq.~\eqref{eq:wT} in Eq.~\eqref{eq:LST}.  

Eq.~\eqref{eq:LST} shows that a ferroelectric transition, $\ve_0 \to \infty$, implies a softening of one of the TO modes. In STO, as the temperature is lowered, the static dielectric function steadily increases from $300$ to $2\times 10^4$ following a Curie-Weiss behavior $\ve_0(T)\propto (T-T_0)^{-1}$ that signals a ferroelectric instability at around $T_{0}\sim 36$ K [see Fig.~\ref{fig:FE_diag}(d)]. However, the enhancement levels off below $40$ K and the dielectric constant saturates, such that STO remains paraelectric down to the lowest temperatures. This behavior has been assigned to a crossover from a classical paraelectric to a quantum paraelectric state, in which the ordered state is suppressed by quantum fluctuations~\cite{Muller1979,Rowley2014}. 
In agreement with the lattice dynamical theory of Cochran~\cite{Cochran1960}, the softening of the TO phonon mode related to the incipient ferroelectricity [see Fig.~\ref{fig:FE_diag}(b)] has been extensively reported in infrared spectroscopy~\cite{Barker1962,Spitzer1962}, neutron scattering~\cite{Cowley1962,Cowley1964,Yamada1969}, and Raman experiments~\cite{bauerle1980soft,Vogt1981,Yamanaka2000}. The temperature dependence of this so-called ferroelectric (FE) soft mode is shown in Fig.~\ref{fig:FE_diag} (e), going from $11$ meV at room temperature down to $1$ meV at around $5$ K, but never reaching condensation on cooling. 

STO can nevertheless be tuned into the ferroelectric phase in various ways, as discussed in Sec. \ref{Sec:intro}. The first experiments used uniaxial strain applied along the pseudocubic directions $[100]$ and $[110]$
to tune STO across the ferroelectric transition ~\cite{Burke1971,Uwe1976}. More recently, room-temperature ferroelectricity in STO films was obtained by exploiting the epitaxial strain imposed by the substrate Ref.~\cite{Haeni2004}. 
The fact that other Ti-based perovskites display ferroelectricity motivated the use of chemical substitution on the cation site to tune STO across the FE transition.
Substitution of Sr with very low concentrations of a $Z$ cation Sr$_{1-x}Z_x$TiO$_3$ such as Ca ($x_c=0.0018$)~\cite{Bednorz1984}, Ba ($x_c=0.035$)~\cite{Lemanov1996} or Pb ($x_c=0.002$)~\cite{Lemanov1997} was found to induce ferroelectricity at a critical doping concentration $x_c$. Isotope substitution of oxygen ${}^{16}$O by ${}^{18}$O at $x_c=0.33$ triggers a finite ferroelectric transition temperature as well~\cite{Itoh1999,Wang2001}. 

At $T_{\mathrm{AFD}}=105$ K, STO undergoes a cubic-to-tetragonal structural transition (to space group $I4/mcm$), with a small distortion $c/a=1.00056$~\cite{Lytle1964}. 
The associated zone-corner R-point optical (nonpolar) phonon $R_{25}$ becomes soft at the transition~\cite{Fleury1968Soft}. Across it, the positions of the Sr and Ti atoms remain fixed, while the oxygen octahedra rotate about one of the cubic axes, with opposite rotation in adjacent cells, as illustrated in Fig.~\ref{fig:FE_diag}(c). For this reason, this is known as an antiferrodistortive (AFD) transition, and the primitive unit cell is doubled below it. The axis about which the octahedral rotation happens is elongated in the tetragonal phase. Therefore, in unstrained samples, there is domain formation with the three possible orientations of the octahedral rotation about the cubic axes [see Fig.~\ref{fig:intro}(d)].
A polarized Raman study recently mapped these domains in the tetragonal state~\cite{Gray2016}. The presence of domain walls may have an impact on superconductivity, as we discuss in more detail in Sec. \ref{Sec:Perspectives}.

The symmetry-breaking at the AFD transition also reconstructs the phonon spectrum. In particular, the symmetry of the FE soft phonon mode is lowered from the three-dimensional $T_{1u}$ representation to a two-dimensional $E_u$ representation (with displacements perpendicular to $[001]$) and a one-dimensional $A_{2u}$ representation (with displacements along $[001]$)
below $T_{\mathrm{AFD}}$, with the corresponding phonon frequencies splitting as $\w_{E_u}<\w_{A_{2u}}$ (see Fig.~\ref{fig:FE_diag}(e)). Moreover, new even-parity phonon modes appear at the zone center 
due to the doubling of the unit cell. Because of its finite electron-phonon matrix element, the $A_{1g}$ soft phonon has also been proposed as a source of attraction for superconductivity~\cite{Appel1969}.

\subsection{Electronic Structure}

\begin{figure}
 \begin{center}
    \includegraphics[width=\linewidth]{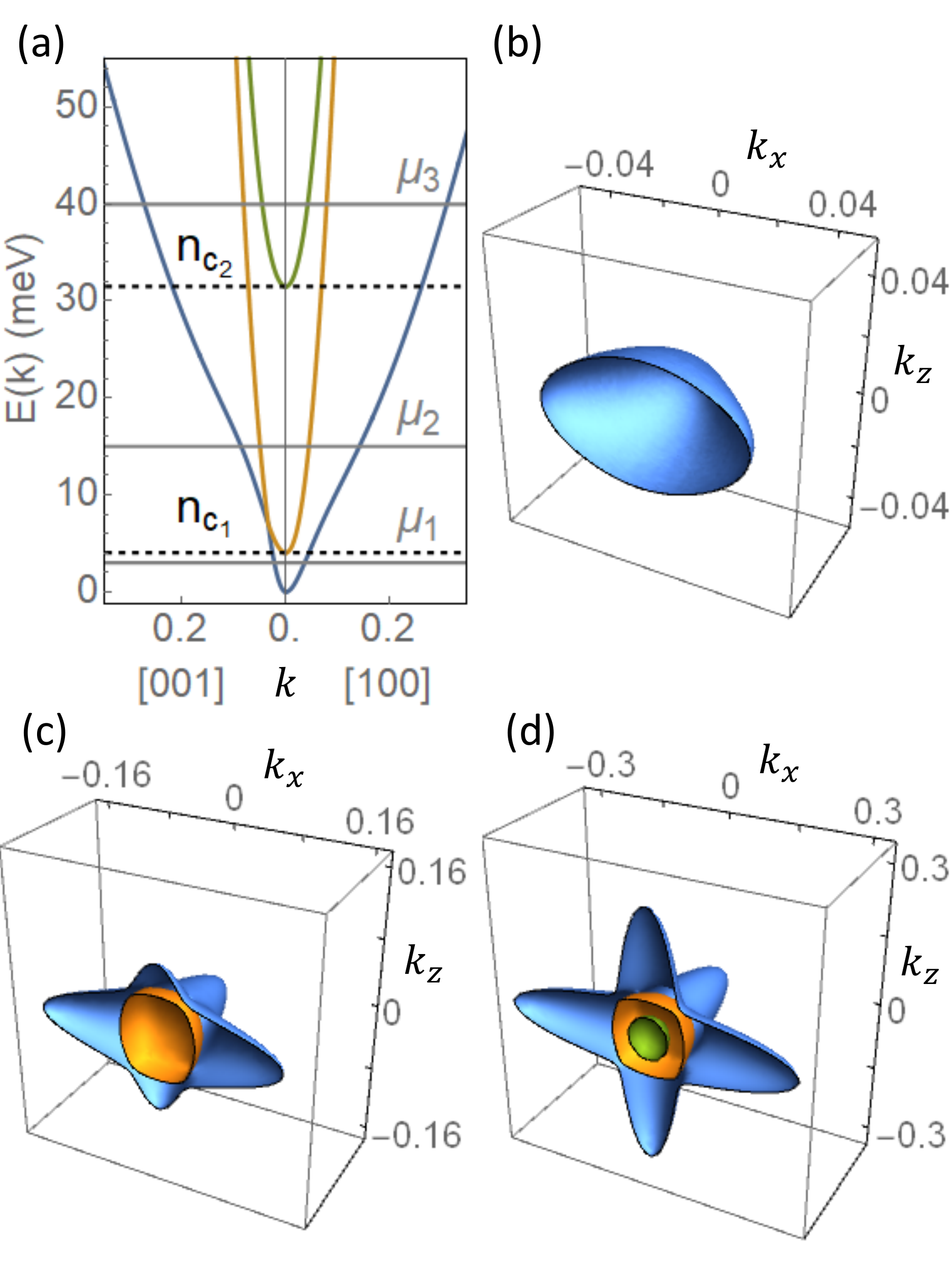}
 \end{center}
\caption{(a) Band dispersions for the tight-binding model of Eq.~\eqref{eq:TB-model} with parameters specified in Table~\ref{tab:numbers}. They reproduce the DFT results very well at low energies~\cite{vanderMarel2011}. The dashed lines $n_{ci}$ indicate the concentrations where Lifshitz transitions take place. The Fermi surfaces for three chemical potential values $\mu_i$ indicated in (a) (gray lines) are shown in (b) $\mu_1=3$ meV (one-band), (c) $\mu_2=15$ meV (two-bands) and (d) $\mu_3=40$ meV (three-bands). In all panels $k$ is in units of $\frac{\pi}{a}$. }
 \label{fig:TB-model}
\end{figure}

In terms of its electronic properties, STO is a band insulator with a $3$ eV gap between the occupied oxygen $2p$ bands and the unoccupied Ti $3d$ $t_{2g}$ bands~\cite{Shanthi1998}. In the low-temperature tetragonal phase, DFT band structure calculations~\cite{Mattheiss1972,vanderMarel2011,Zhong2013} find three electron bands around the zone center with $4$ meV and $27$ meV energy splittings at $\vec k=0$ [see Fig.~\ref{fig:TB-model}(a)].

The low-energy band structure can be successfully described by a minimal tight-binding model~\cite{Bistritzer2011,vanderMarel2011,Zhong2013}
$ \label{eq:tight_binding}
H = \sum_{\bs k}\psi_{\bs k}^\dag \mc H(\bs k)\psi_{\bs k}\,,
$
where the spinor  $\psi_{\bs k}^\dag$ is expressed in the $t_{2g}$ basis ($yz\downarrow,xz\downarrow,xy\uparrow$). 
$\mc H(\bs k)$ includes the triply degenerate $t_{2g}$ orbitals, the atomic spin-orbit coupling  term $\mathcal{H}_\xi$, and a tetragonal crystal field term $\mathcal{H}_\Delta$, 
\begin{align}
\label{eq:TB-model}
\mathcal{H}(\bs k)&=\mathcal{H}_0(\bs k)+\mathcal{H}_\xi+\mathcal{H}_\Delta\\\nonumber
&=\begin{pmatrix}
\epsilon_X(\vec k) & 0 & 0\\
0 & \epsilon_Y(\vec k) & 0\\
0 & 0 & \epsilon_Z(\vec k)
\end{pmatrix}+
\frac{\xi}{2}\begin{pmatrix}
 0&-i & 1\\
 i&0&i\\
 1&-i&0
\end{pmatrix}\\\nonumber
&\quad+\Delta \begin{pmatrix}
1&0&0\\
0&1&0\\
0&0&-2
\end{pmatrix}
\end{align}
where, 

\begin{equation}\label{eq:aux_TB-model}
\epsilon_i(\vec k)=\epsilon_0+4 t_1 \sum_{j\neq i} \sin^2\left(\frac{k_j}{2}\right)+4 t_2\sin^2\left(\frac{k_i}{2}\right)-\mu.
\end{equation}
In the cubic phase, spin-orbit coupling lifts the sixfold degeneracy of the $d_{yz}$, $d_{xz}$ and $d_{xy}$ orbitals into a quartet $\Gamma_8^+$ ($j=3/2$) with energy  $-\xi/2$ and a doublet $\Gamma_7^+$ ($j=1/2$) with energy $\xi$. The tetragonal crystal field that onsets below $T_{\mathrm{AFD}}=105$ K further breaks the four-fold degeneracy of the lower state $\Gamma_8^+$ into two-fold degenerate states. Fig.~\ref{fig:TB-model}(a) shows the resulting band dispersion fitted to the DFT band structure with parameters specified in Table~\ref{tab:numbers}. Note the substantial anisotropy of the lowest band, by comparing its dispersion along the $[001]$ and $[100]$ directions. The strong directional dependence of this band is also manifested in the shape of the Fermi surface, as shown by the blue surface in Figs.~\ref{fig:TB-model}(b)-(d), which correspond to chemical potential values of $\mu_1=3$ meV (one-band filled), $\mu_2=15$ meV (two-bands filled) and $\mu_3=40$ meV (three-bands filled), respectively. The middle (orange) and upper (green) bands, on the other hand, are more isotropic and display quasi-spherical Fermi surfaces. 

A finite density of mobile electrons can be introduced in STO by n-type doping with Nb, La, or oxygen vacancies. This leads to a very dilute metallic state with densities as low as $8\times 10^{15}$cm$^{-3}$~\cite{Spinelli2010}, which however displays a sharp Fermi surface as seen by quantum oscillations for densities of the order of $5\times10^{17}$cm$^{-3}$~\cite{Lin2014}. The main features of the DFT electronic band structure [Fig.~\ref{fig:TB-model}(a)] agree with detailed Shubnikov-de Haas measurements of O deficient and Nb-doped STO \cite{Lin20133}. In particular,
the multiple quantum oscillation frequencies observed experimentally as the carrier concentration increases signal the onset of two Lifshitz transitions at $n_{c1}=1.2\times 10^{18}$ cm$^{-3}$ and $n_{c2}=1.6\times 10^{20}$ cm$^{-3}$. At these Lifshitz transitions, the chemical potential moves up in energy, such that the closest electron-like band sinks below the Fermi level (see Fig. \ref{fig:TB-model}). The impact of these Lifshitz transitions on the superconducting state will be further discussed in Sec. \ref{Sec:Phenomenology}. 

\begin{figure}
 \begin{center}
    \includegraphics[width=0.75\linewidth]{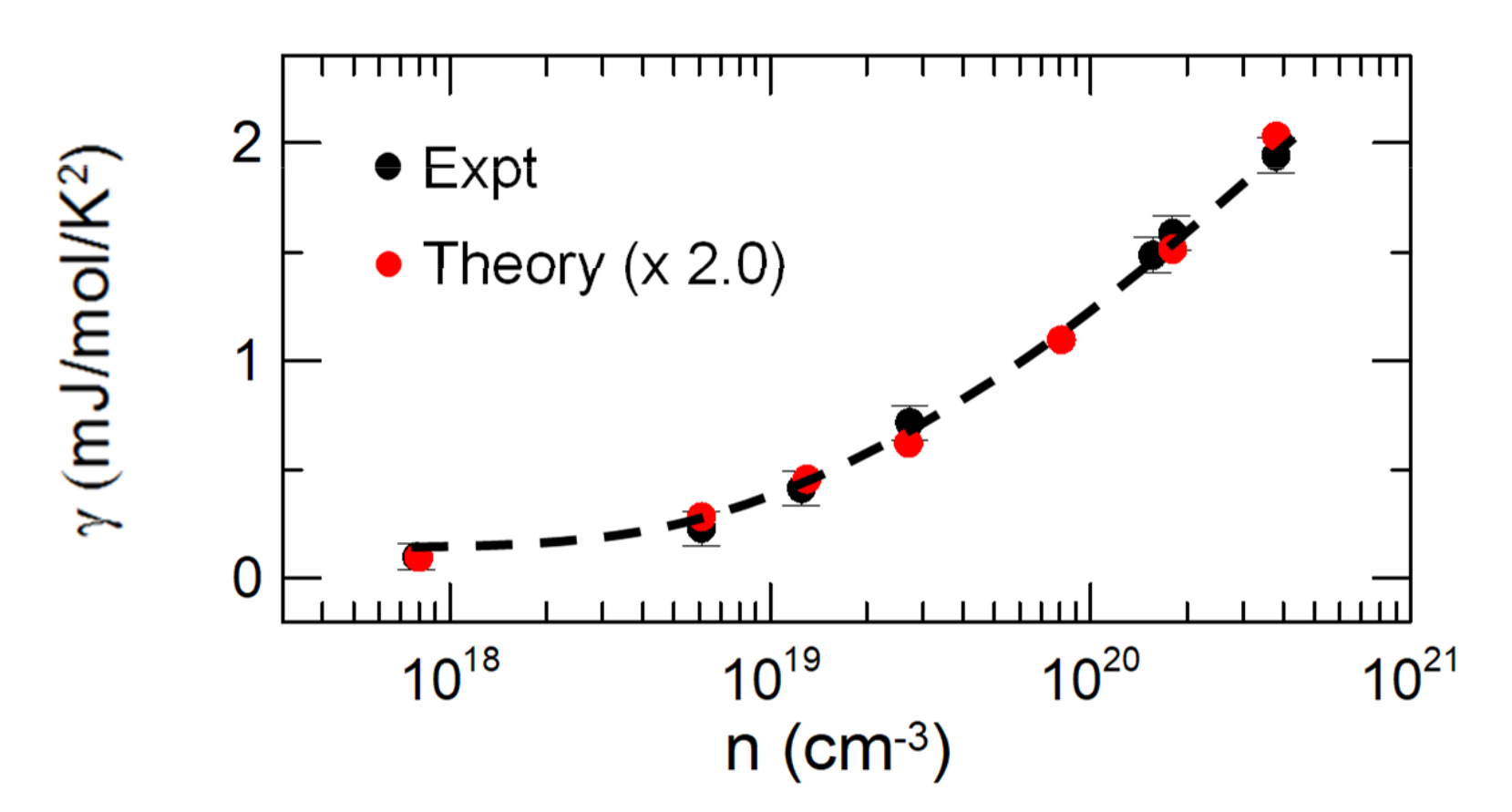}
 \end{center}
\caption{Electronic density dependence of the Sommerfeld coefficient $\gamma$ measured from specific heat experiments (black points) and calculated using the tight-binding model in Fig.~\ref{fig:TB-model}(a) (red points), as reported in Ref.~\cite{McCalla2019}. Note that the theoretical values must be multiplied by a density-independent mass-enhancement factor of $2$ to agree with the data. Figure reproduced with permission from Ref. \cite{McCalla2019}. Copyright 2019 by the American Physical Society}
 \label{fig:sommerfeld}
\end{figure}

As the carrier concentration increases, quantum oscillations and specific heat measurements show that the effective electron mass $m^*$ increases from about $2m_e$ at low doping to about $5 m_e$ at large doping, where $m_e$ is the mass of the electron \cite{Lin20133,McCalla2019}. This doping-dependent enhancement of $m^*$ is not a consequence of electronic interactions, but rather a result of the anisotropic character of the bands. Indeed, as shown in Fig.~\ref{fig:sommerfeld}, the ratio between the experimentally extracted $m^*$ and the theoretical $m^*_{\mathrm{th}}$ 
determined from the tight-binding model fit to DFT calculations [Eq.~\eqref{eq:TB-model}] remains $m^*/m^*_{\mathrm{th}}\approx2$ for the entire range of doping concentrations investigated \cite{vanderMarel2008,vanderMarel2011,McCalla2019}.

\section{Microscopic Pairing Mechanisms}\label{Sec:Pairing}
In section~\ref{Sec:intro} we discussed why the observation of superconductivity in STO is surprising, and briefly 
described the theoretical scenarios that have been proposed to understand this puzzling observation. In this section we delve deeper into some of these theoretical ideas, focusing on their technical details and on the remaining issues they raise.

\subsection{The dynamically screened Coulomb interaction}
The electron-electron interaction in STO is well understood, and is given by the dynamically screened Coulomb interaction:
\be\label{eq:Vc}
V_C(\w,q) = {4\pi e^2 \over \ve(\w,q)q^2}
\ee
The key point is that there are two sources of screening, one arising from the polar phonons [with corresponding dielectric constant $\ve_p(\w,q)$] and another one from the electronic liquid [with corresponding dielectric constant $\ve_e(\w,q)$]:
\be
\ve(\w,q) = \ve_p(\w,q)+\ve_e(\w,q)\,
\ee
As described in Section \ref{Sec:NormalState}, the phonon contribution comes from three dynamical modes (corresponding to three pairs of longitudinal and transverse optical phonon branches) and can be approximated by Eq.~\eqref{eq:LST}. The contribution from the electronic subsystem is given, within the random-phase approximation, by:
\be
\ve_e(\w,q) = -{4\pi e^2 \over q^2}\Pi_e(\w,q)
\ee
where $\Pi_e(\w,q)$ is the electronic polarization bubble.

The main question is: can the interaction in Eq. \eqref{eq:Vc} explain the superconducting state of STO? In the standard case of a longitudinal phonon-mediated interaction with $\w_L \ll \e_F$, the pairing problem can be solved via the standard Eliashberg equations~\cite{Gurevich1962}. In diagrammatic terms, it corresponds to neglecting vertex corrections (which is justified by Migdal theorem) and computing the Nambu self-energy self-consistently via the rainbow diagram. It also neglects the feedback of the fermions on the phonons mediating the interaction, which is also justified by Migdal theorem. In the STO case, it is not obvious that these are the only diagrams that must be considered, particularly for energies larger than $\e_F$~\cite{Grabowski1984}. One possible way to proceed is to write down the Eliashberg-like equations and then afterwards check how/if the contributions from other diagrams affect the outcome. Assuming isotropic $s$-wave pairing within a single band, the Eliashberg equations are given by:   
\begin{widetext}
\begin{align}\label{eqs:Eliashberg-density}
&\phi(i\w_n, k)=-{T \over \nu (2\pi)^3}\sum_{{|n'|<n_c}}\int_0 ^{k_c} {dp \,p^2}\,{\phi(i\w_{n'}, p)\over D(i\w_{n'}, p)}\G(i\w_n-i\w_{n'}, k, p)\\
&\tilde{\xi}(i\w_n, k)= \xi_{ k} + {T \over \nu (2\pi)^3}\sum_{{|n'|<n_c}}\int_0 ^{k_c} {dp \,p^2}\,{\tilde{\xi}(i\w_{n'}, k')\over D(i\w_{n'}, p)}\G(i\w_n-i\w_{n'}, k,  p)\\
&Z(i\w_n, k)=1-{T \over \nu \w_n(2\pi)^3}\sum_{{|n'|<n_c}}\int_0 ^{k_c} {dp \,p^2}\,{\w_{n'}Z(i\w_{n'}, p)\over D(i\w_{n'}, p)}\G(i\w_n-i\w_{n'}, k,  p)
\end{align}
with bare pairing vertex:
\be\label{eq:Gamma}
\G(i\w_n, k,  p) = {\nu \over 4\pi}\oint d\W_{\bs p} V_C(i\w_n, \bs k-  \bs p)\,.
\ee
\end{widetext}

\begin{figure}
 \begin{center}
    \includegraphics[width=0.8\linewidth]{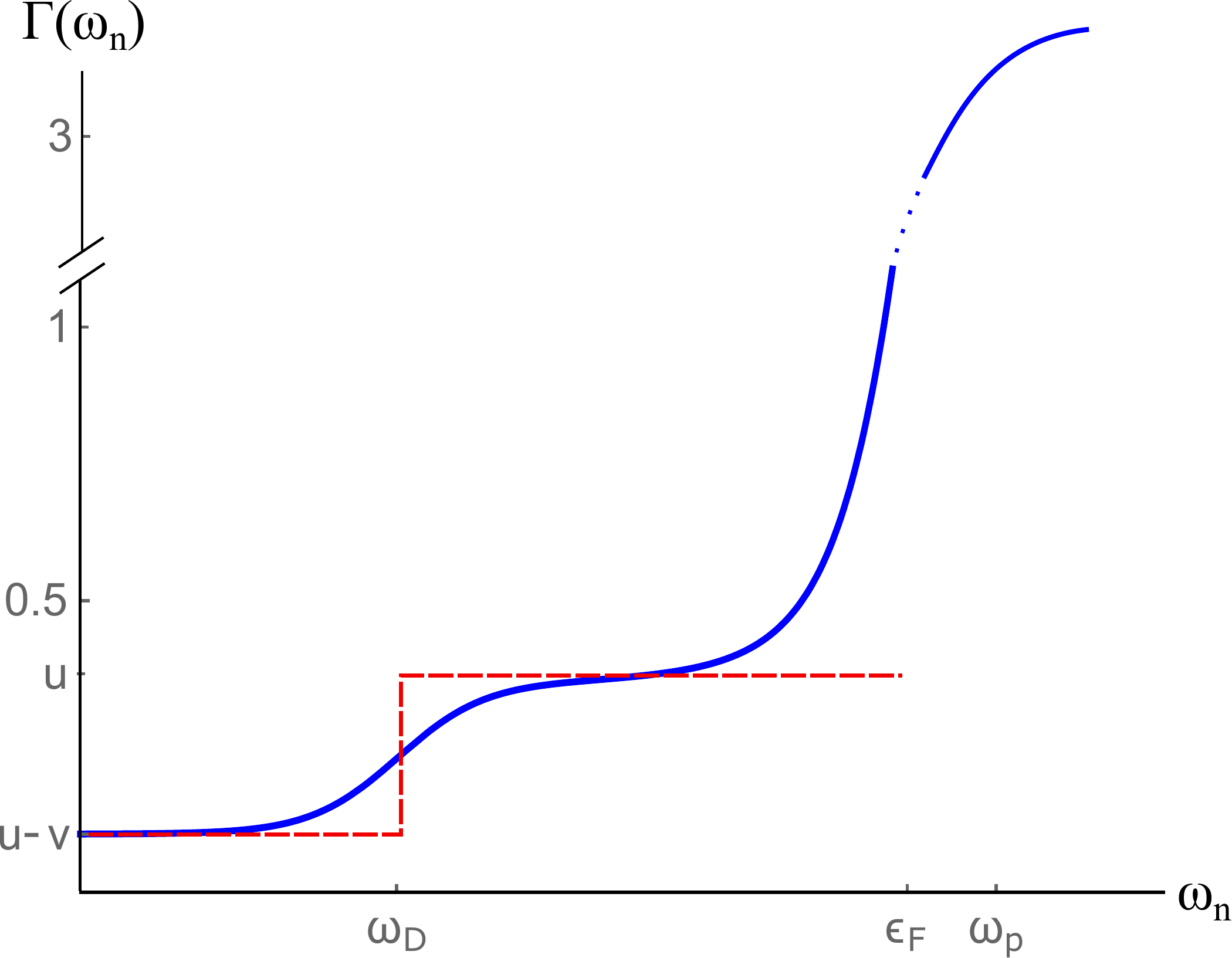}
 \end{center}
\caption{Typical pairing interaction vertex in a metal with dimensionless density $r_s$ of order 1 and with weak electron-phonon coupling. The blue curve represents the sum of the dynamically screened Coulomb and longitudinal-phonon mediated interactions. The red (dashed) curve represents an approximated interaction, see Ref.~\cite{Morel1962}, which consists of a constant repulsive contribution $u$ that is partially reduced below $\w_D$ by an attractive contribution $-v$. Note that $u>v>0$}
 \label{fig:int_metal}
\end{figure}

Here, $\W_{\bs k}$ is the solid angle of $\bs k$, $\nu$ is the density of states at the Fermi level for a parabolic band, $T$ is the temperature, $\w_n = \pi T( 2n+1)$ are Fermionic Matsubara frequencies, $D(i\w_n, k) = \left[\w_n Z(i\w_n, k) \right]^2 + \tilde{\xi}^2(i\w_n, k)+\phi^2(i\w_n, k)$, $\xi_{ k} = \e_{ k}-\mu$, with $\mu(T=0)=\e_F$ and $\e_{ k}$ denoting the parabolic dispersion. $k_c$ and $n_c$ cut off the momentum integral and Matsubara frequency sum, respectively. In addition to these three equations, the chemical potential $\mu$ must also be determined self consistently to fix the total density~\cite{phan2019kohnluttinger}. The meaning of the three unknown quantities is the usual one: $Z$ denotes the imaginary part of the normal component of the self-energy; $\tilde{\xi}$ is the renormalized dispersion due to the real part of the normal component of the self-energy; and $\phi$, proportional to the gap, is the anomalous component of the self-energy.

Before attempting to solve Eq. \eqref{eqs:Eliashberg-density}, important insight can be gained from the frequency dependence of the pairing vertex in Eq.~\eqref{eq:Gamma}. Indeed, a standard approximation employed in solving the Eliashberg equations consists of neglecting the dependence of the gap on the momentum $k$, reducing the linearized problem to a matrix equation in Matsubara space (the appropriateness of this approximation will be discussed below).
First, we note that the pairing vertex is repulsive at all Matsubara frequencies, including the static limit $\w_n \to 0$. 

\begin{figure*}
 \begin{center}
    \includegraphics[width=1\linewidth]{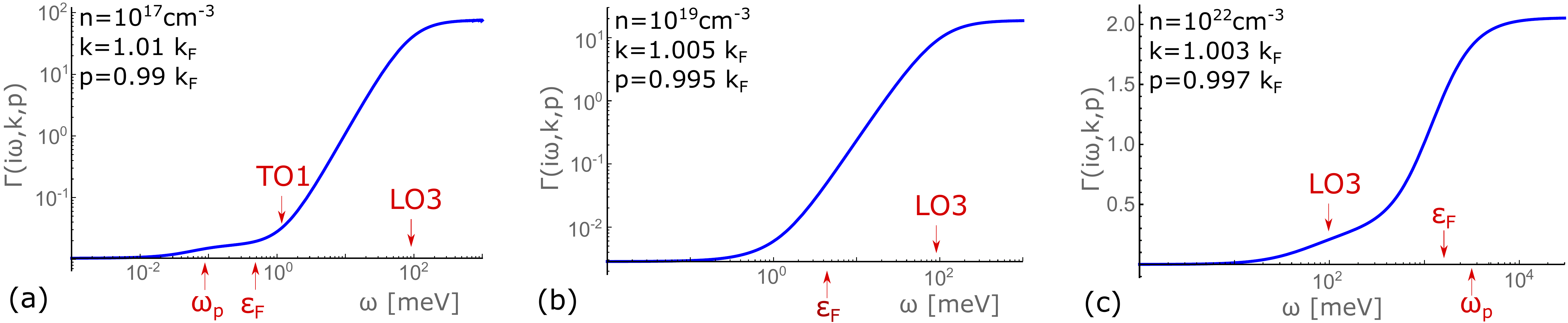}
 \end{center}
\caption{The frequency dependence of the pairing vertex $\Gamma$ of Eq.~\eqref{eq:Gamma} for fixed momenta and carrier concentrations $n = 10^{17}$ cm$^{-3}$ (a), $n = 10^{19}$ cm$^{-3}$ (b), and $n = 10^{22}$ cm$^{-3}$(c). The plasmon frequency $\w_p$, the Fermi energy $\e_F$, and the longitudinal (LO) and transverse (TO) optical phonon frequencies are indicated by red arrows.}
 \label{fig:int}
\end{figure*}

To understand how pairing can emerge from such a purely repulsive interaction, we quickly revisit the standard case of electron-phonon superconductivity with $\w_D \ll \e_F$. The frequency-dependent pairing vertex in this case can be roughly approximated by a step function, $\Gamma(\w_n) = u - v \, \theta\left(\w_D - \w_n\right)$, with $u>v>0$ and a cutoff of the order of $\e_F$ (see Fig.~\ref{fig:int_metal}). Here, $\theta(x)$ is the Heaviside step function. In this simplified BCS-like model, the interaction is always repulsive, but the repulsion $u$ is suppressed due to the contribution from an attractive part $-v$ below the phonon frequency $w_D$. The gap equation admits a piece-wise solution with the gap changing sign at the frequency $\w_D$, yielding a BCS-like gap of the form $\Delta \sim \exp{\left[-1/(v-u^*)\right]}$ (see, for instance, \cite{Coleman2015}). Here, $u^* < u$ is the so-called Coulomb pseudo-potential, which is nothing but the repulsion $u$ suppressed by particle-particle excitations, $u^* = u/\left( 1 + u \ln{\frac{\e_F}{\w_D}}\right)$. We refer to this suppression of the Coulomb repulsion as the Tolmachev-Anderson-Morel mechanism \cite{Tolmachev1958,Morel1962}. Thus, even though $v<u$, a pairing state is possible as long as $v>u^*$. 

Going now back to the pairing vertex in Eq.~\eqref{eq:Gamma}, the infinite frequency limit gives the bare Coulomb repulsion. As the frequency is reduced, the dynamical modes (plasmons and longitudinal polar phonons) come into affect and reduce the repulsion, which reflects the screening. In analogy to the analysis above, each such reduction of the repulsion is essentially an attractive contribution to the overall pairing interaction. Due to the density dependence of the electronic screening, however, the frequency profile of the pairing vertex depends strongly on density. 

In Fig.~\ref{fig:int} we plot the pairing vertex, Eq.~\eqref{eq:Gamma}, in STO, as a function of Matsubara frequency for three different values of the density: $n = 10^{17}$ cm$^{-3}$, $n = 10^{19}$ cm$^{-3}$ and $n = 10^{22}$ cm$^{-3}$. Important frequencies are marked by red arrows. Note that here, for simplicity, we have considered a single parabolic band with effective mass $m^* = 2m_e$~\cite{vanderMarel2011}. 

Focusing on Fig. \ref{fig:int}(a), corresponding to the very low density limit $n=10^{17}$ cm$^{-3}$, the frequency dependent pairing vertex clearly displays two distinct steps.
The first high-frequency step, starting near the LO3 frequency ($\w_{L,3}\sim 100$ meV, see Table~\ref{tab:numbers}), is the contribution from the longitudinal optical modes of STO. Although there are essentially three optical modes in this range, their frequencies are close to each other, such that they become indistinguishable on this plot and behave as a single resonance. The reduction is of the order of $10^2$, reflecting a very strong coupling to these modes (which is essentially a Fr\"ohlich coupling).
This is not surprising since, above these modes' frequencies, the dielectric constant is $\mc O(1)$, whereas below this frequency range, it becomes $\mc O(10^4)$. Thus, this reduction reflects the dimensionless gas parameter $r_s$ (defined in section ~\ref{sec:dynamical_coulomb}) taken above the resonances (with $\ve_\infty$), which is roughly 50. 
At a lower frequency, a small additional reduction (of order 0.01) appears at the plasmon frequency $\w_p$. Notice that the Fermi energy is higher than the plasmon mode but lower than the optical phonons. 

In the intermediate regime with $n=10^{19}$ cm$^{-3}$, shown in Fig. \ref{fig:int}(b), the plasmon frequency is higher than the Fermi energy and hybridizes with the optical phonon modes. As a result, there is essentially a single step of the pairing vertex in which the repulsion is suppressed. The Fermi energy lies somewhere between the step and zero frequency. 

Finally, in the high density regime with $n=10^{22}$ cm$^{-3}$, displayed in Fig. \ref{fig:int}(c), the plasmon frequency reemerges as a well defined mode above the optical modes. As a result, there are again two distinguishable steps resulting from the attractive contributions of the pairing interaction. Here, the LO3 mode leads to a reduction of the repulsion of order $0.25$, much smaller than the case in Fig. \ref{fig:int}(a). The reduction in the overall pairing interaction as function of doping is clear from the comparison between the y-axis scales of the figure. This reflects the drop in the bare value of the dimensionless density parameter $r_s$. 
Note that the plots in Fig. \ref{fig:int} consider specific values of momenta. As we will discuss later, the pairing vertex also depends on $k$ and $p$.

Strictly speaking, the GLF theory \cite{Gurevich1962} discussed in Sec. \ref{sec:dynamical_coulomb} applies to the situation plotted in panel (c), in which the Fermi energy is clearly higher than the frequency of the optical mode, where the smaller step in the pairing vertex takes place. The problem with applying the GLF theory appears as the density is lowered, and the systems moves to the anti-adiabatic limit of a Fermi energy smaller than the phonon frequency.   Concomitantly, the coupling to the optical modes grows with decreasing density and eventually becomes much larger than 1.

\begin{figure}
 \begin{center}
    \includegraphics[width=0.8\linewidth]{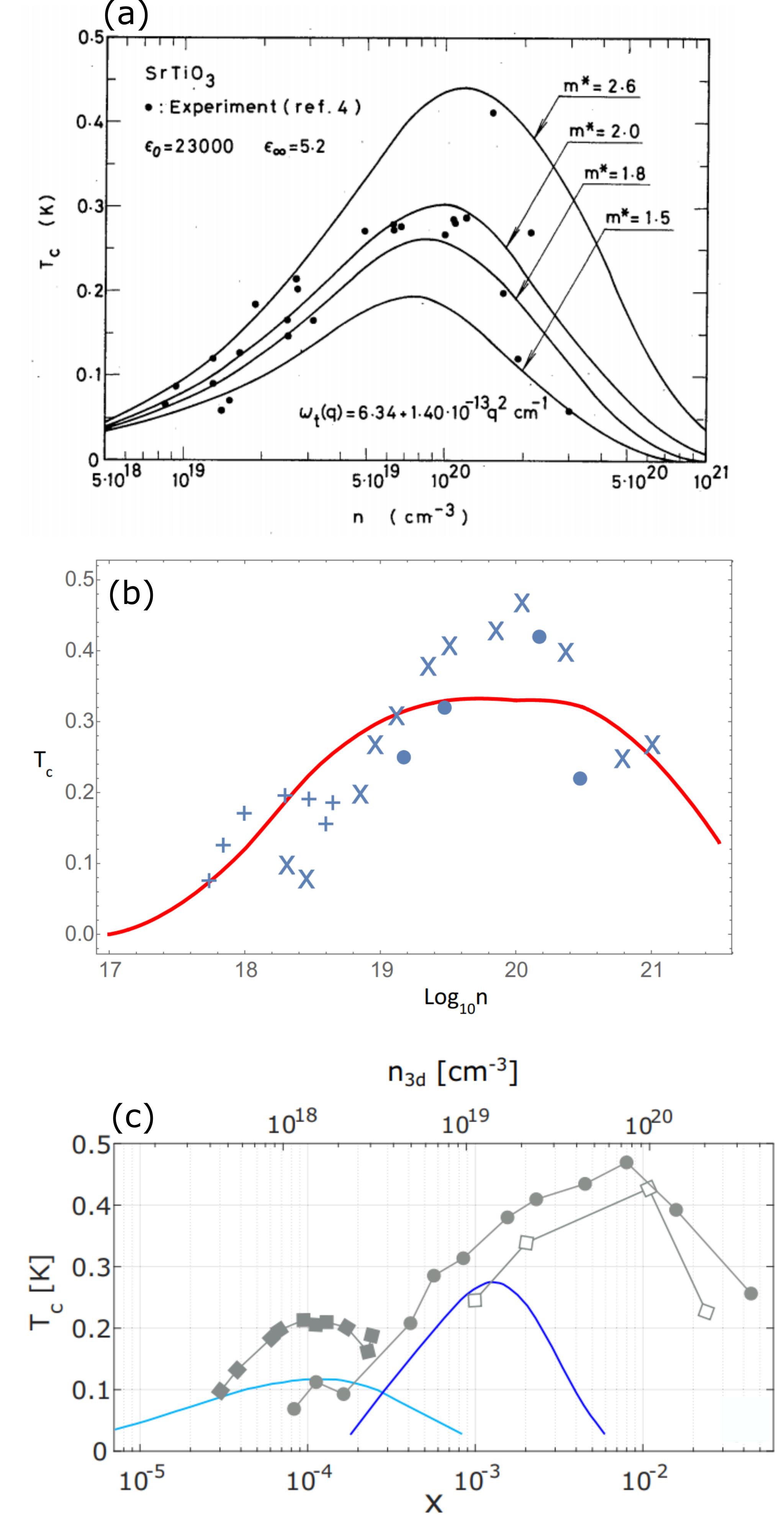}
 \end{center}
\caption{Comparison between different theoretical predictions for $T_c$ as function of the carrier density $n$ in STO, compared to experimental data. All calculations involve approximate solutions of the Eliashberg equations \eqref{eqs:Eliashberg-density}. Panel (a) refers to \cite{takada1980theory}; panel (b), to \cite{wolfle2018superconductivty}; and panel (c), to \cite{ruhman2016superconductivity}. Panel (a) reproduced with permission from Ref. \cite{takada1980theory}. Copyright 1980 by the Physical Society of Japan. Panel (b) reproduced with permission from Ref. \cite{wolfle2018superconductivty}. Copyright 2018 by the American Physical Society. Panel (c) reproduced with permission from Ref. \cite{ruhman2016superconductivity}. Copyright 2016 by the American Physical Society. }
 \label{fig:Tc}
\end{figure}

Notwithstanding these issues, a variant of the Eliashberg equations~\eqref{eqs:Eliashberg-density}, based on the Kirzhnits-Maksimov-Khomskii (KMK) approximation~\cite{Kirzhnits1973}, was solved in 1980 by Takada~\cite{takada1980theory} using the entire frequency range of the vertex~\eqref{eq:Gamma} while ignoring the momentum dependence. The $T_c$ calculated from this approach agreed well with the experimental data of Ref. \cite{Schooley1964}, as shown in Fig.~\ref{fig:Tc}(a). More recently, in Ref. \cite{Klimin2019}, a somewhat related calculation using the KMK approximation including the full non-parabolic band structure Eq.~\eqref{eq:TB-model} was performed by Klimin \etal to explain the isotope effect~\cite{stucky2016isotope}. 
Rowley \etal~\cite{rowley2018superconductivity} also used such an approximation to explain the carrier concentration and pressure dependence of $T_c$ in Nb-doped STO. The pairing interaction in these works is mediated by hybrid longitudinal optical modes, which couple free carriers and ions.

Going beyond the KMK approximation, W\"olfle and Balatsky solved the frequency-dependent Eliashberg gap equation, assuming a momentum-independent gap function \cite{wolfle2018superconductivty,wolfle2019reply}. Their result for $T_c(n)$, shown in Fig.~\ref{fig:Tc}(b), agrees well with the experimental data also. To obtain $T_c$, Ref. \cite{wolfle2018superconductivty,wolfle2019reply} used for the Matsubara frequency cutoff $\w_c$ the energy beyond which quasi-particles are no longer well-defined, i.e. frequencies for which the imaginary part of the normal self-energy exceeds the quasi-particle energy. This results in a cutoff larger than the Fermi energy. Ref. \cite{wolfle2018superconductivty} argued that the Eliashberg equations remain valid for energies up to the cutoff due to the fact that the coupling remains weak at this energy scale. If the coupling constant is indeed small, vertex corrections can be safely neglected and the absence of Migdal theorem -- issue (i) in Sec. \ref{Sec:Issues} -- is no longer a problem. A different point of view was put forward by Ruhman and Lee \cite{ruhman2019comment}, who argued that additional diagrams beyond those included in the standard Eliashberg equations must be considered for energies larger than the Fermi energy. They also objected to the value of the cutoff used in Ref. \cite{wolfle2018superconductivty}, noting further that due to the strong dependence of the pairing vertex with frequency in the regime above $\e_F$, the choice of cutoff crucially affects the value of $T_c$. This ongoing debate highlights the richness of the problem, and begs for further investigations in this direction.

A different approach to the Eliashberg equations \eqref{eqs:Eliashberg-density} was taken by Ruhman and Lee in Ref. \cite{ruhman2016superconductivity}, focusing specifically on the very low density limit $n\sim 10^{17}$ cm$^{-3}$. They argued that, in this dilute regime, even though the longitudinal phonon frequency is much larger than $\e_F$, there is another bosonic mode whose frequency remains lower than the Fermi energy: the plasmon [see Fig.~\ref{fig:int} (a)]. They contended that, because $\w_p < \e_F$, the plasmon mode can provide the pairing mechanism, and the approximations employed in the standard Eliashberg formalism are well justified. As a result, the interaction in Eq.~\eqref{eq:Vc} was approximated by a single plasmon pole supplemented by a phenomenological parameter $\eta$ to reduce the high-frequency repulsion:
\be
V_C(i\w_n,q) = {4\pi e^2 \over \ve_0 q^2} \left[\eta - {\w_p^2 \over \w_n^2 + \w_p^2} \right]
\ee
Solution of the Eliashberg equations showed that the coupling to this mode is too small to lead to a sizable transition temperature. To allow for a reasonable $T_c$, Ref. \cite{ruhman2016superconductivity} considered $\ve_0$ as an a additional fitting parameter. A good agreement with the experimental data was obtained for $\ve_0\sim 10^3$, as shown in Fig.~\ref{fig:Tc}(c). The fact that this theory needs two additional adjustable parameters ($\eta$ and $\ve_0$) to explain the data suggests that the coupling to the plasmon mode alone is likely not sufficient to account for the superconductivity of STO in the low-density limit. Furthermore, it is not clear that the feedback effect of the fermions on the plasmon propagator can be neglected.

Leaving aside for a moment the question of whether the diagrams included in the standard Eliashberg equations \eqref{eqs:Eliashberg-density} are justified in the case of STO, a rather unexplored issue is about the appropriateness of neglecting the momentum dependence of the gap and self-energy functions, as it was done in most of the attempted solutions of those equations described above. The fact that the gap is likely $s$-wave (more on this in Sec. \ref{Sec:Phenomenology}) only justifies integrating out the dependence on the momentum coordinates tangential to the Fermi surface. As for the perpendicular momentum component, the standard approximation within the Migdal-Eliashberg theory is to replace it by the Fermi momentum. Furthermore, the renormalization of the electronic dispersion by the real part of the self-energy ($\tilde{\xi}$ in the Eliashberg equations) is also neglected within the standard approach. While these approximations are very reasonable when $\w_L \ll \e_F$, in dilute STO this condition is clearly not satisfied.

\begin{figure}
 \begin{center}
    \includegraphics[width=0.6\linewidth]{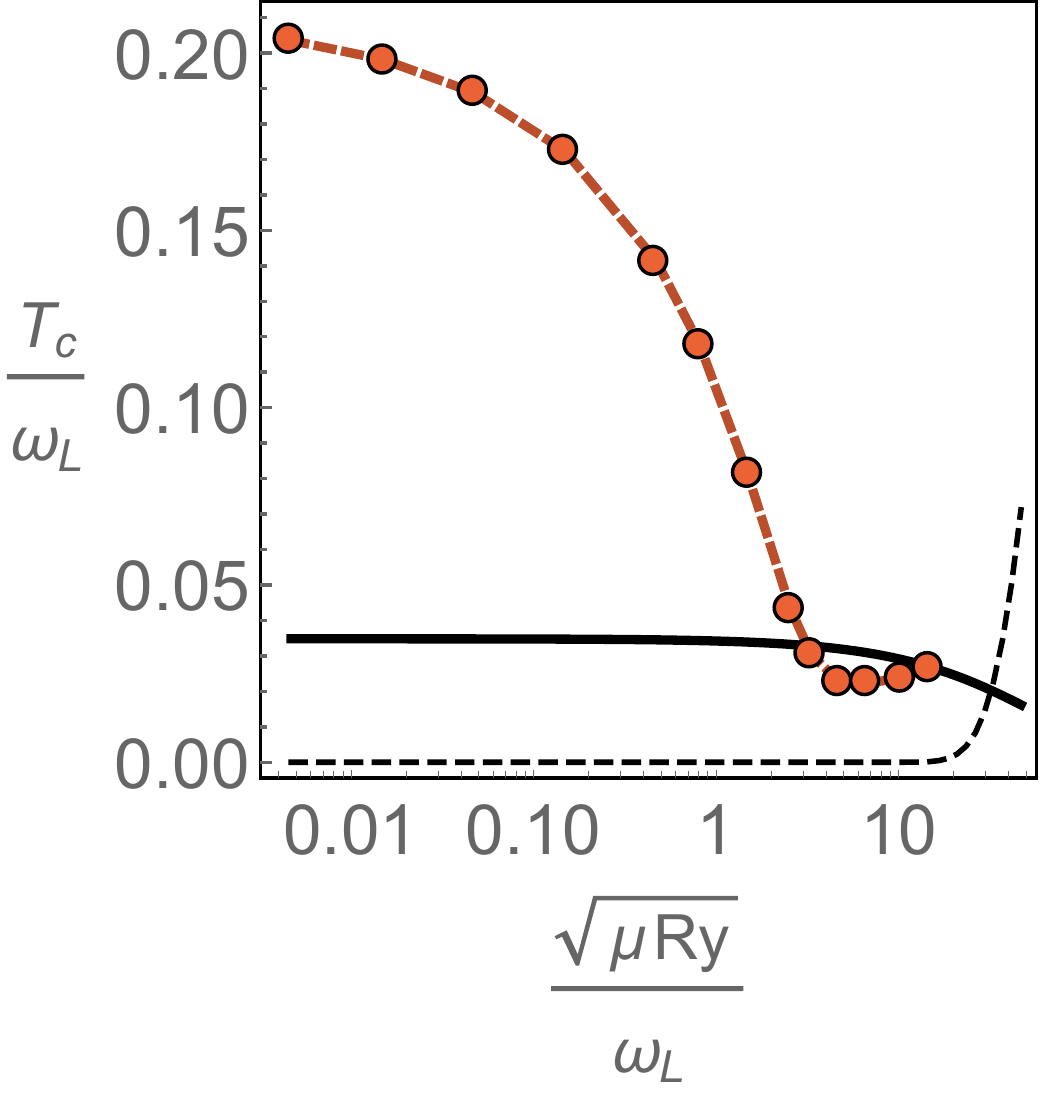}
 \end{center}
\caption{Transition temperature $T_c$ of the Bardeen-Pines-like model of Eq. (\ref{eq:BP}), as reported in Ref. \cite{Gastiasoro2019}, as function of the bare chemical potential $\mu$ (evaluated at $T_c$) for a fixed phonon frequency $\w_L$. The black (red) curve is the solution without (with) the momentum dependence of the pairing interaction included. The dashed line is the conventional BCS expression. Here, Ry is the Rydberg energy.}

 \label{fig:Tc_momentum}
\end{figure}

In Ref. \cite{Gastiasoro2019}, Gastiasoro, Chubukov, and Fernandes investigated the impact of the momentum dependence of the pairing interaction in a much simpler model than Eq. \eqref{eq:Vc}, solving the full set of Eliashberg equations (\ref{eqs:Eliashberg-density}). In particular, they considered the attractive part of a Bardeen-Pines-like electron-phonon interaction:

\be\label{eq:BP}
V(i\w_n,q) = -\frac{4\pi e^2}{q^2 + \kappa^2}\left( \frac{\w_L^2}{\w_n^2 + \w_L^2} \right)
\ee
where $\kappa$ is the Thomas-Fermi screening momentum. The authors found that, as the Fermi energy goes to zero in the extreme dilute limit, contributions from states far away from the Fermi level become increasingly more important to the pairing problem. As a result, they argued that the perpendicular momentum component dependence of the Eliashberg equations cannot be neglected in this limit. 
Fig. \ref{fig:Tc_momentum} illustrates how $T_c$ is affected by such contributions. In particular, an enhanced $T_c$ was found in the limit of $\e_F \to 0$, displaying a polynomial dependence on the phonon frequency, $T_c \sim \w_L \left( \mathrm{Ry}/ \w_L\right)^{1/5}$, where Ry is the Rydberg energy. It remains an open question how important these effects are in the case of the more complicated interaction \eqref{eq:Vc}.

We finish this section by briefly mentioning a completely different approach for the pairing mechanism in STO, in which the pairing interaction $V_C(i\w_n,q)$ is attractive in the static limit, $\w_n \to 0$. This contrasts to the pairing interaction in Eq.~\eqref{eq:Vc}, which remains repulsive for all frequencies. Mechanisms that could promote a local attractive static interaction have been proposed by Gor'kov \cite{gor2015phonon,Gorkov2017} and by Geballe \cite{kivelson2019physics}. Gor'kov proposed that localized polar impurities can lead to overscreening of the Coulomb repulsion, rendering the bare interaction attractive \cite{Gorkov2017}. It is interesting to point out that such moments have been recently seen in experiments~\cite{wang2019charge}. Geballe's proposal is that oxygen vacancies can act as negative-$U$ centers and thus induce local attractive interaction, via a mechanism that is similar to what was proposed to explain superconductivity in Tl-doped PbTe \cite{Matsushita2006}.

Finally, before moving on to other mechanisms, we note that in the interaction Eq.~\eqref{eq:Vc} we have considered the bosonic modes (e.g. plasmons and optical phonons) within the random-phase approximation. In particular, the self-energy renormalization of these bosons was neglected. In the standard case, where $\e_F$ is much greater than the entire bosonic frequency range and the Migdal criterion holds, this is a good approximation. 
However, the effects of this renormalization should be taken into account when the Fermi energy is comparable to the bosonic mode frequency, which has not been considered so far in the STO literature.

\subsection{Pairing from quantum critical ferroelectric fluctuations}
As discussed in Sec. \ref{Sec:NormalState}, it is believed that quantum fluctuations are strong in STO, as they prevent the onset of long-range ferroelectric order while stabilizing a quantum paraelectric state. The possibility that these ferroelectric quantum fluctuations can be responsible for the pairing mechanism in STO has led to a considerable amount of experimental ~\cite{Rowley2014,stucky2016isotope,rischau2017ferroelectric,herrera2018strainengineered,tomioka2019enhanced,rowley2018superconductivity,van2019possible,wang2019charge} and theoretical~\cite{Edge2015,kedem2016unusual,wolfle2018superconductivty,wolfle2019reply,ruhman2019comment,kedem2018novel,Kanasugi2018spin,Arce-Gamboa2018quantum,Kozii2019Superconductivity}  works. Experimentally, it is generally observed that $T_c$ is enhanced when STO is tuned closer to the putative ferroelectric quantum critical point (i.e. a $T=0$ continuous phase transition) \cite{Rowley2014}, which can be accomplished via $^{18}$O substitution \cite{stucky2016isotope,tomioka2019enhanced,wang2019charge}, strain \cite{herrera2018strainengineered,Harter2019,Ahadi2019}, ``negative" pressure \cite{rowley2018superconductivity}, or Ca doping \cite{rischau2017ferroelectric}. Theoretically, the exchange of critical ferroelectric fluctuations provide an alternative mechanism to the dynamically screened Coulomb interaction that also promotes long-range attractive interactions in STO. Moreover, it places STO inside a larger class of unconventional superconductors in which quantum critical fluctuations have been proposed to be responsible for Cooper pairing (see e.g. \cite{Metlitski2015}).

Although the phenomenological theory of displacive ferroelectricity dates back to many decades ago, see e.g. \cite{Kwok1966,Rechester1971,khmelnit1971low,Vaks1968}, only more recently the role of quantum fluctuations \cite{Roussev2001Quantum,Coleman2009,Simons2010} and the coupling to gapless electronic states in a metal have been considered ~\cite{fu2015parity,Kozii2015,wolfle2018superconductivty,Kozii2019Superconductivity}.
Across this displacive-type structural transition [see Fig.~\ref{fig:FE_diag}(b)], the crystal structure loses inversion symmetry. Due to the polarity of the ions in the unit cell, the breaking of inversion symmetry also induces a dipolar electric moment in pristine samples. In doped samples, the free charge carriers screen the dipolar fields, implying that macroscopic ferroelectricity is absent. Nevertheless, one still uses the term metallic ferroelectric to refer to a metal that undergoes a phase transition that can locally induce a dipole moment. Indeed, as we will see, the long-range dipolar fields promote electronic interactions that have important effects.  

The displacive structural transition is characterized by a gapless optical phonon mode described by the action ~\cite{Roussev2001Quantum,ruhman2019comment}
\be
\mc S_u = {1\over 2}\sum_{\bs q} \chi^{-1}_{ij}(\bs q, \w) u_{i}u_j + \mc O (u^4)
\ee
where the phonon propagator is given by:
\be \label{eq:chi_phonon}
\chi^{-1}_{ij}(\bs q, \w) = D_{ij}(\bs q) -\w^2 \delta_{ij}
\ee
with the static component:
\begin{align}
D_{ij}(\bs q) &= \w_T^2\d_{ij} + c_T^2\left( q^2\d_{ij}- q_i  q_j\right) \nonumber \\ 
&+\left[ c_L^2+{\W_I^2 \over  q^2} \right] q_i  q_j+\a q_i^2 \d_{ij} \label{eq:A(q)}
\end{align}

Here $c_T^2$ and $c_L^2$ are the transverse and longitudinal velocities, respectively. $\w_T$ is the TO frequency, which controls the distance to the quantum critical point (QCP), such that $\w_T \to 0$ at the QCP. $\a$ represents the anisotropic cubic crystal field terms.

\begin{figure}
\begin{centering}
\includegraphics[width=1\columnwidth]{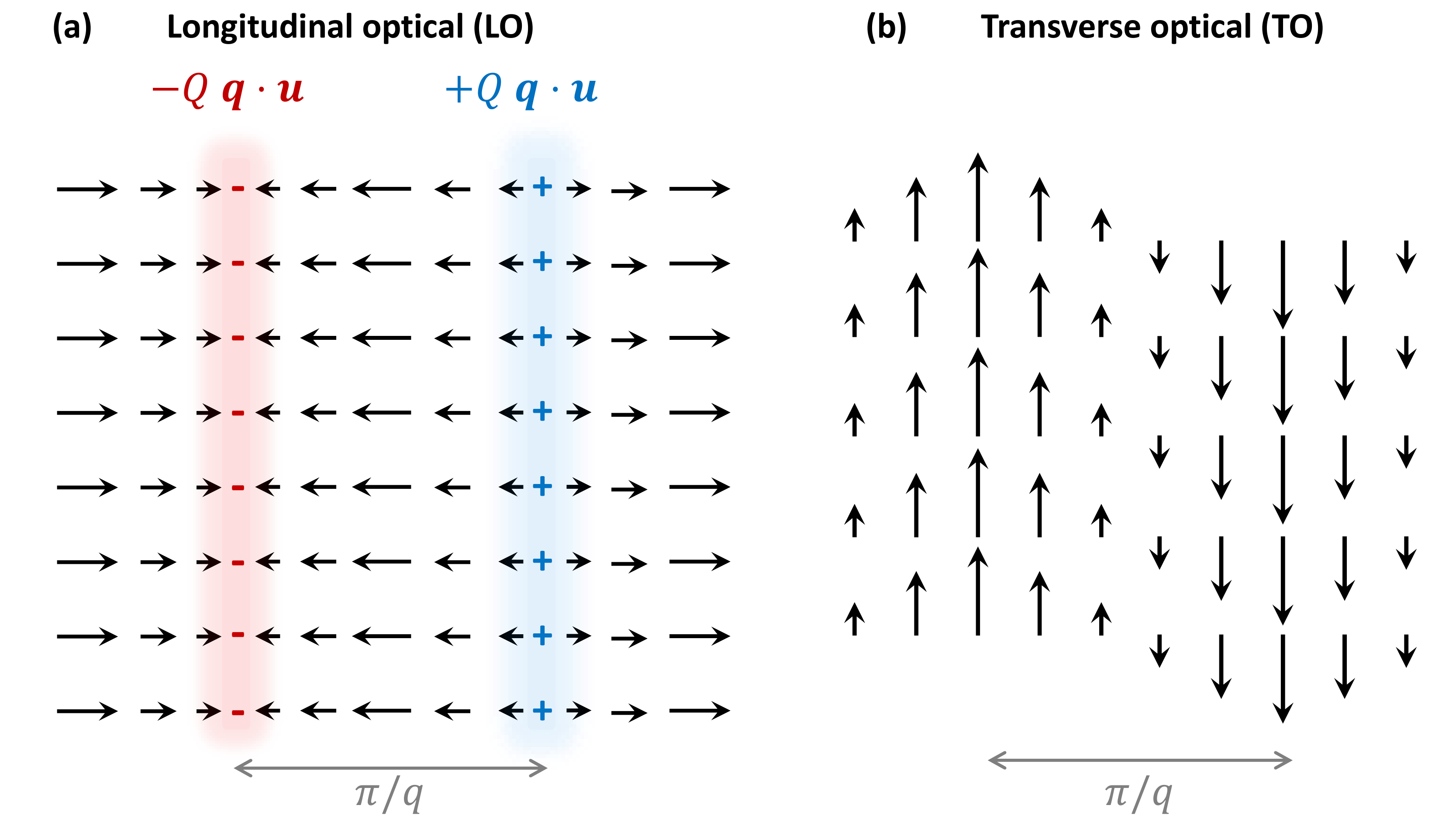}
\par\end{centering}
\caption{Schematic plot of polarization waves. (a) The longitudinal polarization wave causes a charge modulation. Similar to the plasma wave in a charged liquid, the long-range Coulomb force leads to a gap $\W_I$. Consequently, the LO phonon mode is gapped at the transition point. (b) The transverse polarization wave, on the other hand, does not induce such a charge gradient and is therefore free to become soft at the transition.  \label{fig:LO-TO}}
\end{figure}

The term proportional to $\W_I^2$ in Eq. (\ref{eq:A(q)}) represents the Coulomb energy generated by an LO deformation [see Fig.~\ref{fig:LO-TO} (a)], where $\W_I = \sqrt{4\pi Q^2 \rho_I / \ve_\infty M}$ is the ionic plasma frequency. Here, $\rho_I$, $M$ and $Q$ are the ionic density, mass and charge, respectively. The TO deformation, on the other hand, is decoupled from the ionic charge given by $Q\nabla \cdot \bs u$ [see Fig.~\ref{fig:LO-TO}(b)]. In the limit of $q\to 0$, the $\W_I$ term leads to a finite gap between the LO and TO branches (the LO-TO splitting~\cite{mahan2013many}). As a result, the LO frequency, which is given by $\w_L = \sqrt{\w_T^2 + \W_I^2}$, remains gapped at the ferroelectric QCP, where $\w_T \to 0$. 

The connection to ferroelectricity follows from the fact that a finite displacement vector $\bs u$ [Fig.~\ref{fig:FE_diag}(b)] generates a finite polarization $\bs P$, i.e. $\bs P \propto \bs u$. Thus, even if a macroscopic polarization is suppressed by screening in a metal, the coupling of the elastic field $\bs u$ to the electronic degrees of freedom remains. One could then attempt to use the Eliashberg formalism in Eq. (\ref{eqs:Eliashberg-density}) to compute the pairing instability by replacing $V_c(\bs q, \w)$ in the pairing vertex in Eq. (\ref{eq:Gamma}) by the bosonic propagator $\chi_{ij}(\bs q, \w)$ of Eq. (\ref{eq:chi_phonon}). 

Edge \etal \cite{Edge2015} proposed that the coupling between electrons and fluctuations near the ferroelectric QCP can explain the superconducting dome in STO. Instead of the propagator $\chi_{ij}(\bs q, \w)$, they used an effective transverse-field Ising model to describe the ferroelectric QCP and found that the pairing interaction is proportional to $1/\w_T$. Then, because $\w_T$ decreases as doping increases, whereas the density of states increases, a superconducting dome emerges. Their model also predicted an enhancement of $T_c$ due to oxygen isotope substitution ($^{16}$O $\to$ $^{18}$O), as this would essentially decrease $\w_T$ by moving STO closer to the ferroelectric QCP (see also Ref.~\cite{kedem2016unusual}). Note that this prediction contrasts with the conventional isotope effect, by which $T_c$ should be suppressed upon $^{18}$O substitution. Such an anomalous isotope effect was later confirmed experimentally \cite{stucky2016isotope}. 

Two important issues that were not addressed in Ref. \cite{Edge2015} and remain under debate are (i) the impact of the dynamics of the critical fluctuations on the superconducting instability, and (ii) how the electrons couple to the soft TO mode. Point (i) certainly deserves further investigation, particularly because studies of superconductivity mediated by critical fluctuations associated with other QCPs suggest that the dynamical part of the pairing interaction plays a key role \cite{Abanov2003}. In particular, the feedback of the fermions on the bosonic propagator is known to fundamentally alter the bosonic dynamics in the cases of antiferromagnetic and ferromagnetic QCPs. As for point (ii), more recent works have provided further insight into it \cite{wolfle2018superconductivty,ruhman2019comment,wolfle2019reply}. The challenge in coupling electrons to a transverse phonon mode is apparent when one considers the most common electron-phonon gradient coupling: 
\be\label{eq:grad_coupling}
\mc S_{uc}^{\mathrm{ph}} =\sum_{\bs k \bs q}\l_{\mathrm{ph}, \bs q} \left(i\bs q \cdot \bs u_{\bs q}\right) c_{s \bs k+\bs q}^\dag c^{\phantom{\dagger}}_{s \bs k}
\ee
Here, $c_{s \bs k}$ annihilates an electron with spin projection $s$ and momentum $\bs k$ (summation over spin indices is left implicit). For example, the Fr\"ohlich coupling~\cite{frohlich1950xx} falls into this category, with $\l_{\mathrm{ph},q} = {4\pi e Q \over \ve_\infty q^2}$. The point is that Eq.~\eqref{eq:grad_coupling} allows coupling only to the LO branch~\cite{ruhman2019comment}, which does not become soft at the transition. As pointed out by 
 W\"olfle and Balatsky~\cite{wolfle2018superconductivty}, the cubic crystal-field anisotropy, denoted by $\alpha$ in Eq.~\eqref{eq:A(q)}, mixes the LO and TO modes away from high-symmetry directions, resulting in an effective coupling between the electronic density and the ferroelectric modes. However, this mechanism seems to still give a rather small coupling to the TO mode~\cite{ruhman2019comment,wolfle2019reply}. Note that the dipolar coupling to the electron density in Eq.~\eqref{eq:grad_coupling} was also used in other theoretical works that have considered quantum critical ferroelectricity as a pairing mechanism ~\cite{Arce-Gamboa2018quantum,kedem2018novel}.    

An alternative way to directly couple the electronic density to the transverse modes is via the scalar $\bs u^2$, as first pointed out by Ngai~\cite{Ngai}. Indeed, the symmetry constraint on the coupling in Eq. \eqref{eq:grad_coupling} is removed if two-phonon processes are considered. Such a coupling is given by:
\be\label{eq:scalar_coupling}
\mc S_{uc}^{2\mathrm{ph}} =\sum_{\bs k \bs q} \l_{2\mathrm{ph}} \, \bs u^2_{\bs q} \, c_{s, \bs k+\bs q}^\dag c^{\phantom{\dagger}}_{s, \bs k}\phantom{^\dagger}
\ee

The microscopic origin of this coupling are virtual $p-d$ transitions between O and Ti ions (contribution from the Fan-Migdal self-energy term) and the energy shift of the $d$ orbitals of Ti ions (contribution from the Debye-Waller self-energy term)~\cite{Giustino2017}.
This coupling was recently invoked by van der Marel \etal{}~\cite{van2019possible} as a possible mechanism for superconductivity in STO, resurfacing the original proposal by Ngai \cite{Ngai}. They used optical conductivity measurements to estimate the coupling, finding an effective BCS-like coupling constant of $\l_{2\mathrm{ph}}^{\mathrm{BCS}}\approx 0.28$. 
It is quite surprising that the two-phonon processes give such a large BCS-like coupling, but if this is indeed the case, they certainly provide a viable mechanism.

A third possible coupling mechanism emerges in the presence of spin-orbit coupling. In this case, the transverse optical modes are allowed to couple to the electronic density to linear order via~\cite{fu2015parity,Kozii2015,Kanasugi2018spin,Kozii2019Superconductivity,Kanasugi2019} 
\be\label{eq:Rashba_coupling}
\mc S_{uc}^{\mathrm{SOC}} =  \sum_{\bs k \bs q} \l_{\mathrm{SOC}} \left[c_{s, \bs k+{\bs q\over 2}}^\dag \left(\bs k\times \bs \s_{ss'}\right) c^{\phantom{\dagger}}_{s', \bs k-{\bs q\over 2}}\right]\cdot \bs u_{\bs q}
\ee
This coupling is unique in the sense that it remains finite in the limit of $q\to 0$. Therefore, it is potentially a relevant perturbation at the critical point. The impact of the coupling in Eq.~\eqref{eq:Rashba_coupling} on the ferroelectric QCP has not been studied, and very little has been done in connection to superconductivity. 
Recently, Kanasugi \etal{}~\cite{Kanasugi2018spin,Kanasugi2019} studied the interplay between the superconducting state and ferroelectricity by coupling the electronic states to the polar distortion via the Rashba-like coupling of Eq.~\eqref{eq:Rashba_coupling}. As the source of superconductivity, however, they assumed a phenomenological momentum-independent intra-orbital attractive pairing interaction that leads to a uniform $s$-wave state.

Gastiasoro \etal \cite{GastiasoroFE} have explored the weak-coupling superconducting pairing interaction that arises from the coupling \eqref{eq:Rashba_coupling} in the vicinity of the ferroelectric instability. They found that the effective coupling is indeed dominated by the transverse sector, and the leading singlet instability is in the $s$-wave channel. Due to the cubic symmetry of the propagator (\ref{eq:chi_phonon}), the $s$-wave solution does not give an isotropic gap. On the contrary, the gap function found in \cite{GastiasoroFE} acquires an anisotropy that increases as the frequency of the ferroelectric mode goes soft, $\w_{T}\rightarrow 0$. Note that previous works have found that couplings of the form of Eq. \eqref{eq:Rashba_coupling} can also favor triplet pairing \cite{Kozii2015}. 

An important question that deserves further attention is the expected magnitude of the coupling $\l_{\mathrm{SOC}}$ in STO. Ruhman and Lee~\cite{ruhman2016superconductivity} argued that $\l_{\mathrm{SOC}}$ should be of the same order as the weaker of the two inter-orbital hoppings in STO, that is, a few hundreds of meV. A more accurate estimate from first-principle calculations would thus be desirable. However, this may be challenging due to the dense $k$-mesh that is needed to project this coupling onto the Fermi surface states. 

An interesting related problem in which the impact of quantum critical ferroelectric fluctuations on a metal can be studied in a more theoretically controlled manner was investigated by Kozii \etal \cite{Kozii2019Superconductivity}. In particular, motivated by superconductivity in doped Sn$_x$Pb$_{1-x}$Te, they considered the case of a ferroelectric QCP in a Dirac semimetal at charge neutrality. An important difference with respect to the standard metal case is that the valley degrees of freedom allow for a direct coupling between the TO mode and the electronic density. Using a renormalization group approach, they found that the coupling between the TO mode and the electronic density is marginally relevant. They considered the complete low-energy theory including also the Coulomb repulsion, and found a strong enhancement of $T_c$ in the vicinity of the ferroelectric QCP.

\section{Multi-band Superconductivity and Gap Structure}\label{Sec:Phenomenology}

\subsection{The role of inter-band interactions}

The fact that multiple bands of STO cross the Fermi level as doping
increases, according to the tight-binding model of Eq.~\eqref{eq:TB-model} (see also Figs. \ref{fig:intro} and \ref{fig:TB-model}), suggests that multiple superconducting gaps can be present. Given the difficulties in establishing a microscopic model, it is useful to resort to phenomenology to understand the implications of multi-band superconductivity \cite{Fernandes2013,Yanase2013,Chubukov2016,Valentinis2016}. For the three-band case, the linearized gap equations become:

\begin{equation}
\Delta_{i}=-\ln\frac{\Lambda}{T_{c}}\sum_{j=1}^{3}V_{ij}\nu_{j}\Delta_{j}\label{gap}
\end{equation}
where $\Delta_{i}$ is the gap in band $i$, $\nu_{i}$ is the corresponding
density of states, $\Lambda$ is an upper energy cutoff, and the $V_{ij}$
describe intra-band ($i=j$) and inter-band ($i\neq j$) pairing interactions.
Obviously, in view of all the aforementioned issues that plague a
microscopic description of the pairing state in STO, Eq. (\ref{gap})
should not be taken at face value as a statement for the appropriateness
of a BCS-like state in STO, but rather as a useful framework to model
multi-band superconductivity.

A crucial assumption behind Eq. (\ref{gap}) is that the gaps are
isotropic, implying an $s$-wave state. Note however that, as we explain
below, $s$-wave multi-band superconductivity can be very non-trivial.
Experimentally, the full gaps observed in tunneling and optical spectroscopy
measurements are consistent with an $s$-wave state \cite{Binnig1980,Swartz2018,Thiemann2018},
as is the observed ratio of $\Delta/T_{c}$. Recent thermal conductivity
measurements performed below $T_{c}$ by Lin \etal reported a low-temperature dependence
consistent with the Bardeen-Rickayzen-Tewordt behavior typically seen
in conventional $s$-wave superconductors, with no linear-in-$T$
behavior observed at very low temperatures \cite{Lin_nodeless_2015}.

\begin{figure}
\begin{centering}
\includegraphics[width=0.9\columnwidth]{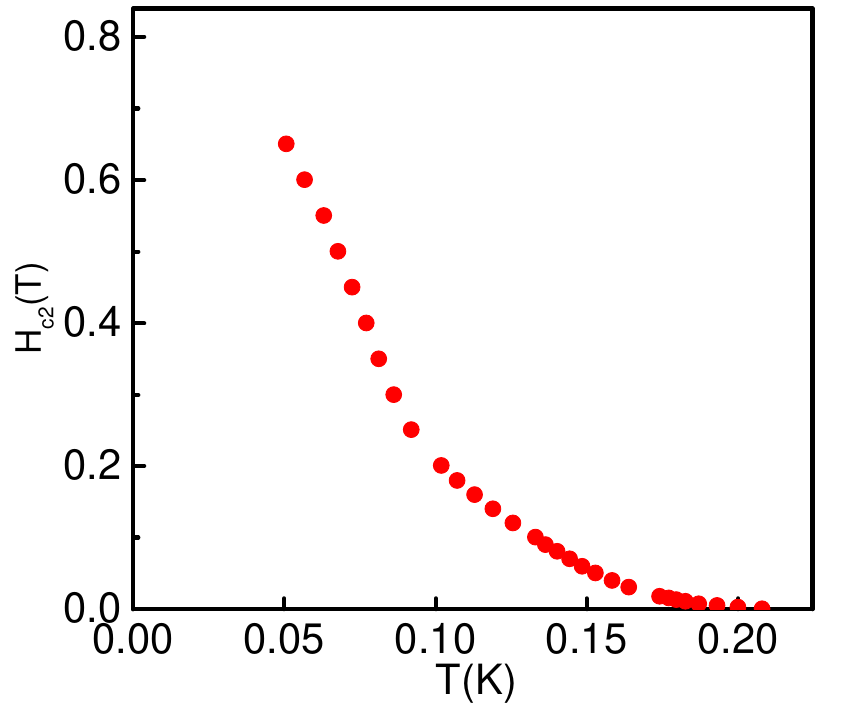}
\par\end{centering}
\caption{Upper $c$-axis critical field $H_{c2}$ (red dots) as function of
temperature for Nd-doped STO. The data is from Ref. \cite{Ayino2018}. The convex shape of the curve near $T_c$ is the behavior expected for a multi-gap superconductor \cite{Edge15}. \label{fig_Hc2}}

\end{figure}

As for direct experimental evidence in favor of multiple superconducting
gaps, the seminal work by Binnig \etal reported two gaps in
the tunneling conductance of Nb-doped STO \cite{Binnig1980}, with
the second gap only emerging at sufficiently high doping concentrations.
Although surface effects may complicate the interpretation of this
result in terms of two bulk gaps, as pointed out recently by Eagles \cite{Eagles2018}, this observation
is consistent with the doping evolution of the band structure of STO, as discussed by Fernandes \etal \cite{Fernandes2013}.
More recently, measurements of the thermal conductivity $\kappa$
and of the critical magnetic field $H_{c2}$ have also provided strong
support for the existence of multi-gap superconductivity \cite{Lin_nodeless_2015,Ayino2018}.
As shown in Fig. \ref{fig_Hc2}, the $H_{c2}(T)$ curve of Nd-doped
STO obtained by Ayino \etal in Ref. \cite{Ayino2018} is convex near $T_{c}$, as
typically seen in dirty multi-gap superconductors. This curvature
was previously predicted theoretically by Edge and Balatsky \cite{Edge15}, and contrasts
to the concave behavior expected for single-gap superconductors.

Theoretically, while the solution of the coupled gap equations (\ref{gap})
requires the knowledge of nine different parameters (three intra-band
interactions $V_{ii}$, three inter-band interactions $V_{i\neq j}$,
and three density of states $\nu_{i}$), some general features of three-band
superconductivity can be inferred after considering a few reasonable
simplifications. The fact that superconductivity is observed in the
single-band regime \cite{Bretz2019} indicates that attractive intra-band
interactions ($V_{ii}<0$) are dominant over inter-band interactions
(which can be repulsive or attractive). We further set all densities
of states to be equal to $\nu$ and simplify Eq. (\ref{gap}) by setting
all intra-band interactions to be the same, $v\equiv-V_{ii}\nu>0$,
and all inter-band interactions to be close in magnitude, $u\equiv V_{12}\nu$,
$u\left(1+\delta_{13}\right)\equiv V_{13}\nu$, and $u\left(1+\delta_{23}\right)\equiv V_{23}\nu$.
The solution of Eq. (\ref{gap}) is then given by $T_{c}=\Lambda\mathrm{e}^{-1/\lambda},$where
$\lambda$ is the largest eigenvalue of:

\begin{equation}
\mathcal{V}=\left(\begin{array}{ccc}
v & -u & -u\left(1+\delta_{13}\right)\\
-u & v & -u\left(1+\delta_{23}\right)\\
-u\left(1+\delta_{13}\right) & -u\left(1+\delta_{23}\right) & v
\end{array}\right)\label{matrix_V}
\end{equation}

The gap structure (i.e. the ratios between the gaps) is given by the
eigenvector $\hat{\Delta}$ of the largest eigenvalue $\lambda$ of
$\mathcal{V}$. While this model is certainly too simplistic to capture
the complexity of STO, it nicely illustrates the non-trivial properties
of multi-gap superconducting states when repulsive interactions are
present -- even if they are not driving the superconducting instability,
i.e. $v\gg|u|$. To see this, consider first the case where all inter-band
interactions are identical, $\delta_{13}=\delta_{23}=0$. When $u$
is also attractive ($u<0$), the largest eigenvalue is $\lambda_{+}=v+2\left|u\right|$
and the corresponding eigenvector, $\hat{\Delta}_{+}=\left(1,\,1,\,1\right)$.
This means that the gap functions are equal and have the same sign
in all three bands. This state, which we dub $s_{+}$, is the extension
of the standard $s$-wave superconducting state to the three-band
case.

\begin{figure}
\begin{centering}
\includegraphics[width=0.9\columnwidth]{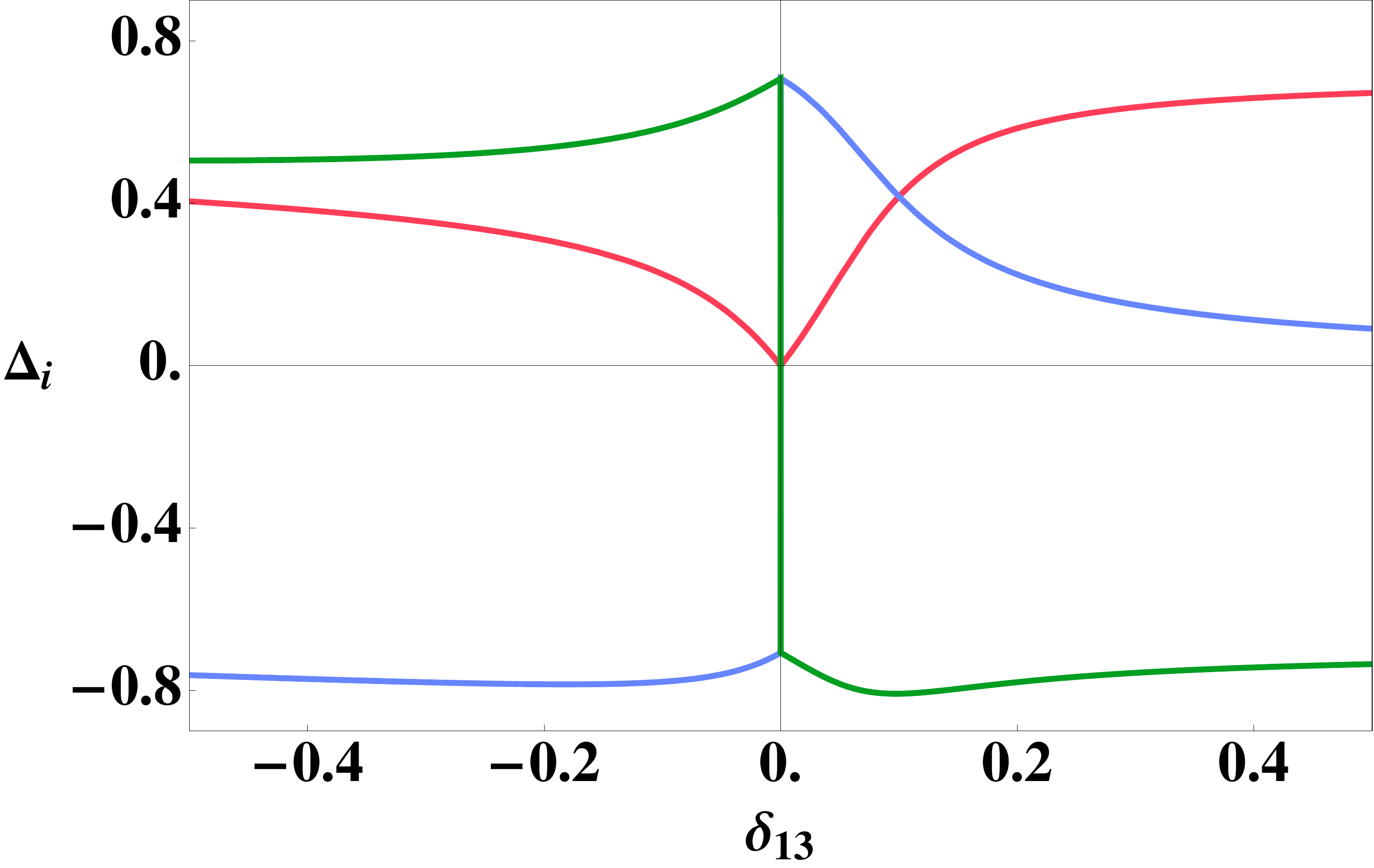}\vspace{0.3in}
\includegraphics[width=0.9\columnwidth]{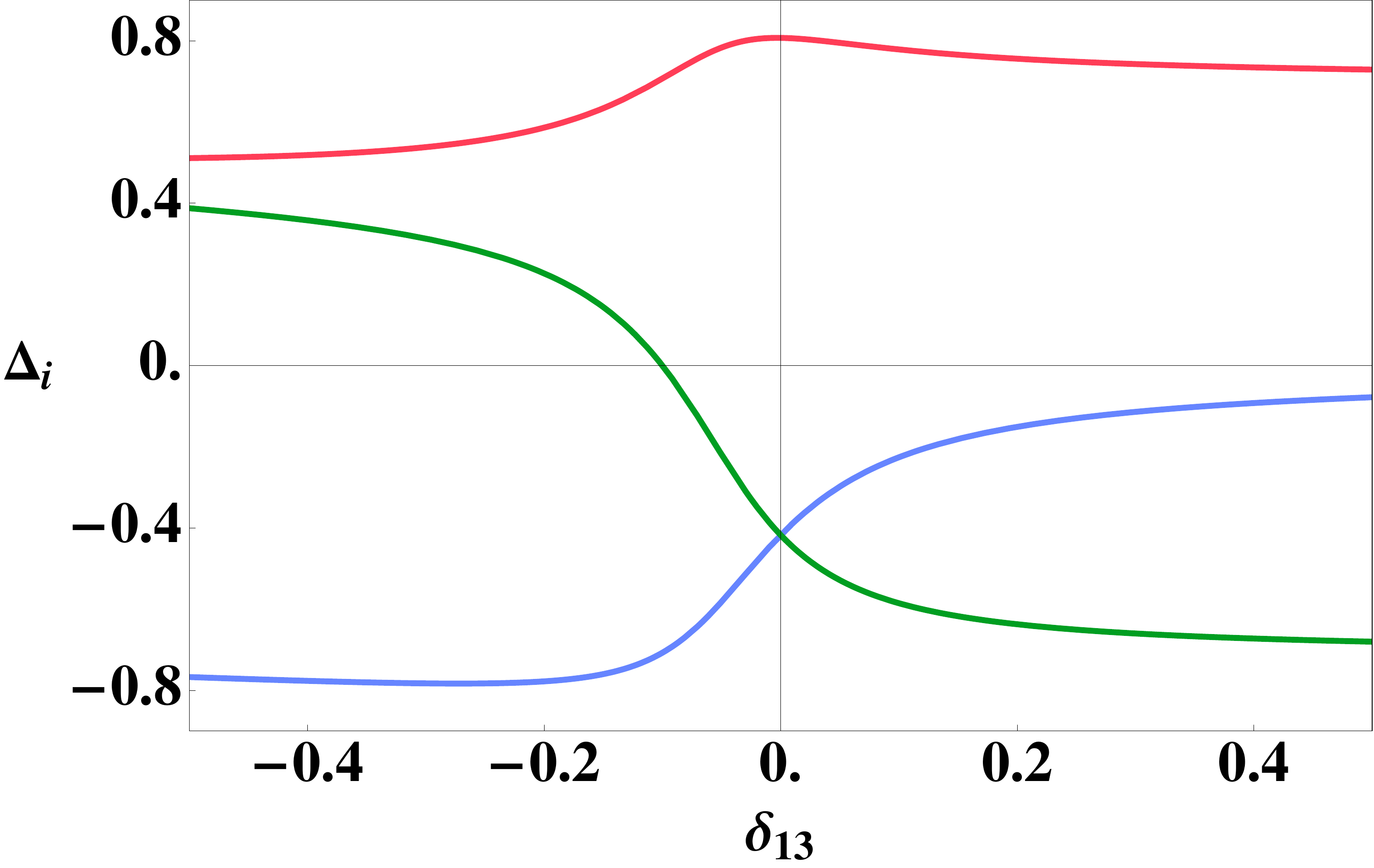}
\par\end{centering}
\caption{Superconducting gaps $\Delta_{i}$ as function of the inter-band interaction
parameter $\delta_{13}$ for $\delta_{23}=0.1$ (upper panel) and $\delta_{23}=-0.1$
(lower panel). The meaning of these parameters is explained in the
text. Red denotes the gap of band $1$; blue, the gap of band $2$;
and green, of band $3$. The gaps are normalized such that $\sum_{i}\Delta_{i}^{2}=1$.
The gap of band $1$ is arbitrarily set to always be positive. \label{fig_3bands}}

\end{figure}

On the other hand, when $u$ is repulsive ($u>0$), the largest eigenvalue
is $\lambda_{-}=v+\left|u\right|$. Interestingly, it is two-fold
degenerate, with eigenvectors $\hat{\Delta}_{-}^{(1)}=\left(1,\,-1,\,0\right)$
and $\hat{\Delta}_{-}^{(2)}=\left(1,\,0,\,-1\right)$. This degeneracy
is a manifestation of the frustration arising from the fact that the
repulsive inter-band interactions impose that the gaps of every pair
of bands should have opposite signs. But because there are three bands,
it is impossible to have a gap configuration in which the signs of
the gaps of every two bands are always opposite. The situation is
analogous to antiferromagnetically-coupled Ising spins on a triangular
lattice. Importantly, this frustration happens even though the inter-band
interaction is sub-leading, i.e. $v\gg|u|$.

This degeneracy is lifted by the small corrections $\delta_{13}$,
$\delta_{23}$ to the inter-band interactions. As shown in Fig. \ref{fig_3bands},
different gap configurations emerge depending on these parameters
(red denotes gap $1$; blue, gap $2$; and green, gap $3$). For $\delta_{23}>0$
(upper panel), the leading eigenvector changes from $\hat{\Delta}_{-}=\left(+,\,+,\,-\right)$
for $\delta_{13}>0$ to $\left(+,\,-,+\right)$ for $\delta_{13}<0$.
For $\delta_{23}<0$ (lower panel), it switches from $\left(+,\,-,\,-\right)$
for $\delta_{13}>\delta_{23}$ to $\left(+,\,-,+\right)$ for $\delta_{13}<\delta_{23}$.
Interestingly, in both cases, there are parameter regimes in which
one of the gaps is much smaller than the other two, which is another
manifestation of the frustration. We dub all these superconducting
states where two pairs of gaps have opposite signs, whereas one pair
of gaps have the same sign, as $s_{-}$.

The richness of the phase diagram of three-band superconductors with
repulsive interactions has been widely discussed in the recent literature,
mostly in the context of iron-based superconductors \cite{Stanev2010},
where inter-band interactions are believed to be the dominant ones
(in contrast to the STO case). Besides suppressing one of the gaps,
the frustration can also lead to more exotic effects, such as the
spontaneous breaking of time-reversal symmetry at a temperature
below $T_{c}$ \cite{Stanev2010}.

The key question is whether any of these interesting effects are present
in STO. To a certain extent, this issue remains little explored, and
further studies are highly desirable. For instance, what would be the 
microscopic mechanism that gives rise to a repulsive inter-band interaction 
but attractive intra-band interaction? The fact that certain experiments
seem to observe just one gap \cite{Thiemann2018,Swartz2018}, or at
most two \cite{Lin_nodeless_2015,Ayino2018}, could be indicative
of the presence of a very small gap that is difficult to observe,
which would be naturally explained by the frustration scenario. Of
course, a more trivial explanation would be that the intra-band interactions
are significantly different for the three bands. Distinguishing between
the $s_{+}$ and $s_{-}$ states is also challenging; a distinct feature
of the latter is the possible existence of a magnetic resonance mode
at the wave-vector that connects the bands whose gaps have opposite
signs \cite{Scalapino2012common}. Given the small wave-vectors and energies
involved, it would be challenging for neutron scattering experiments
to identify such a mode. Collective modes associated with the relative
phase between the gaps -- the so-called Legget modes -- are also
expected to be present below $2\Delta_{0}$, particularly if the intra-band
interactions are the largest ones \cite{Benfatto2013}.

\subsection{Impact of disorder}

Disorder also has a distinct effect in multi-band $s$-wave superconductors,
as compared to the single-band case. In the latter, magnetic impurity
scattering (with scattering rate $\tau^{-1}_{S}$) is pair-breaking,
whereas non-magnetic impurity scattering (with scattering rate $\tau^{-1}_0$)
does not suppress $T_{c}$ globally \cite{Anderson1959,Abrikosov1959}.
In the former, the effect depends on the relative sign between the
gaps ($s_{+}$ or $s_{-}$ state). For a two-band superconductor,
the suppression of $T_{c}$ for weak impurity scattering ($\tau^{-1}\ll T_{c}$)
is given by \cite{Golubov1997}:

\begin{align}
\left(\frac{\Delta T_{c}}{T_{c}}\right)_{s_{+}} & =-\frac{\pi\left(\tau_{S,\mathrm{intra}}^{-1}+\tau_{S,\mathrm{inter}}^{-1}\right)}{4T_{c}}\\
\left(\frac{\Delta T_{c}}{T_{c}}\right)_{s_{-}} & =-\frac{\pi\left(\tau_{S,\mathrm{intra}}^{-1}+\tau_{0,\mathrm{inter}}^{-1}\right)}{4T_{c}}
\end{align}
The key point is that, for both $s_{+}$ and $s_{-}$ states, non-magnetic
intra-band scattering does not affect $T_{c}$, whereas magnetic intra-band
scattering suppresses $T_{c}$. The difference between these two states
resides on the role of inter-band scattering: in the $s_{+}$ case,
only inter-band magnetic scattering suppresses $T_{c}$, whereas in
the $s_{-}$ case, pair-breaking is caused only by inter-band non-magnetic
scattering.

\begin{figure}
\begin{centering}
\includegraphics[width=0.85\columnwidth]{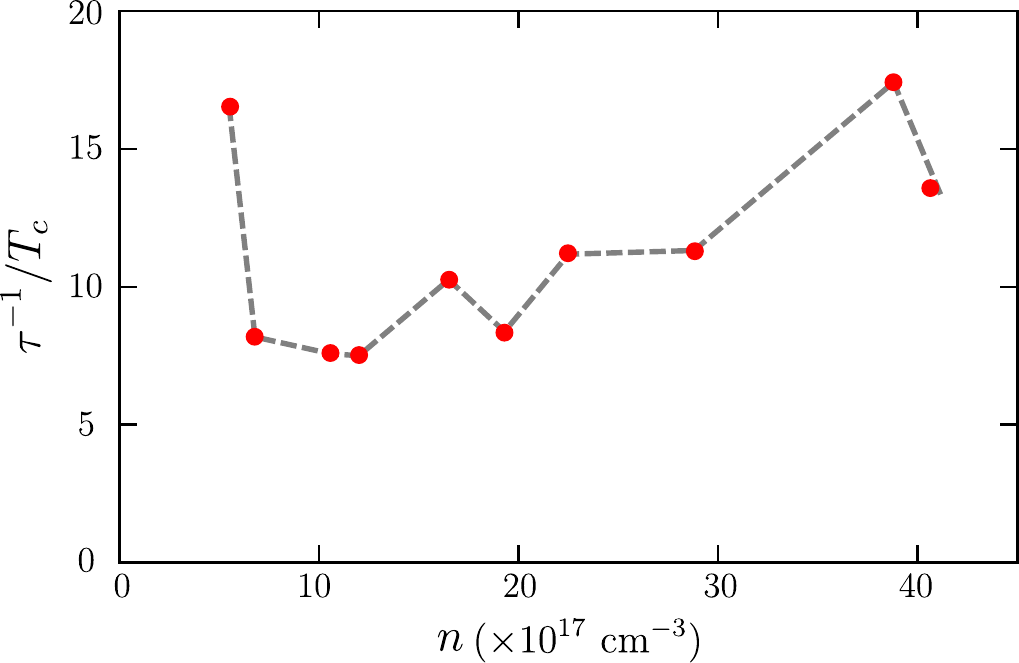}
\par\end{centering}
\caption{Ratio between the total scattering rate $\tau^{-1}$
and the transition temperature $T_{c}$ as function of the carrier
concentration $n$. $\tau^{-1}$ was estimated from
the residual resistivity data of Ref. \cite{Lin20133}. \label{fig_disorder}}

\end{figure}

This distinct response of the $s_{+}$ and $s_{-}$ states to disorder
offers a possible way to experimentally probe the superconducting
ground state of STO (see \cite{Wang2013} for a similar discussion
in the context of iron-pnictide superconductors). The challenge is
on how to experimentally control disorder, since doped STO is intrisically
close to the dirty regime of superconductivity, as pointed out by Collignon \etal \cite{Collignon2017}. Indeed,
as illustrated in Fig. \ref{fig_disorder}, $\tau^{-1}/T_{c}$
can be of the order $10$.

In Ref. \cite{Lin_irraditiona_2015}, Lin \etal employed electron irradiation
to introduce controlled -- and presumably non-magnetic -- disorder,
and $T_{c}$ was found to not change in irradiated samples. Ayino \etal 
reported superconductivity in Nd-doped STO samples in Ref. \cite{Ayino2018}, with $T_{c}$
values comparable to that of Nb-doped STO samples with similar carrier
concentration. Because Nd$^{3+}$ is expected to have a magnetic moment,
this result suggests a weak effect of magnetic impurities on $T_{c}$.
Taken separately, the results of \cite{Lin_irraditiona_2015} and
\cite{Ayino2018} seem to favor an $s_{+}$ and an $s_{-}$ state,
respectively. The crucial obstacle for an unambiguous interpretation
is the difficulty in separating the intra-band and inter-band scattering
contributions. This highlights the need for future studies where both
magnetic and non-magnetic disorder are systematically controlled in
STO.

\begin{figure}
\begin{centering}
\includegraphics[width=0.9\columnwidth]{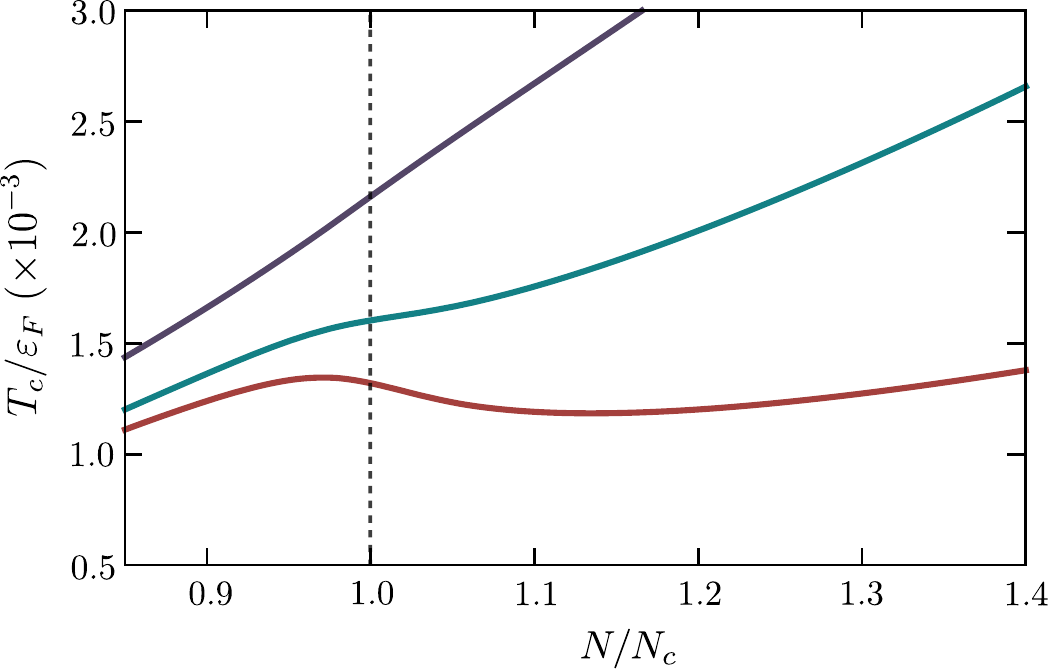}
\par\end{centering}
\caption{Theoretical calculation of $T_{c}$ as function of the carrier concentration
$N$ across the first Lifshitz transition at $N_{c}$, as obtained in Ref. \cite{Trevisan_PRL_2018}.
The dark purple curve shows the case without disorder, and applies for both
$s_{-}$ and $s_{+}$ states. The red (cyan) curve is the disordered
case for the $s_{-}$ ($s_{+}$) state with large non-magnetic impurity scattering. \label{fig_Lifshitz_theory}}
\end{figure}

Impurity scattering was also invoked to explain a peculiar feature
of the superconducting dome of O-deficient STO. As shown in Fig. \ref{fig:Behnia}, Lin \etal found that $T_{c}$ is suppressed across the
first Lifshitz transition \cite{Lin20133}, where the number of bands crossing the
Fermi level increases from 1 to 2. Such a behavior is quite unexpected,
since BCS theory generally predicts, for both $s_{+}$ and $s_{-}$
states, that $T_{c}$ increases across a Lifshitz transition, because
the number of states available to form the superconducting condensate
increases \cite{Fernandes2013,Valentinis2016}. Trevisan \etal \cite{Trevisan_PRL_2018,Trevisan_PRB_2018}
argued that this suppression of $T_{c}$ can be explained if the ground
state is $s_{-}$ and the (non-magnetic) inter-band impurity scattering
is significant. This happens because the pair-breaking effect caused
by inter-band impurity scattering is enhanced once the second band
crosses the Fermi level, overcoming the positive effect on $T_{c}$
caused by the enhancement of the density of states, see Fig. \ref{fig_Lifshitz_theory}.
Such a scenario would thus favor an $s_{-}$ state, and could also
explain why this $T_{c}$ suppression is not always seen in other
STO samples \cite{Bretz2019}, as it depends on the disorder strength.

In the case of repulsive inter-band interactions, increasing disorder
can also change the two-band superconducting ground state from $s_{-}$
to $s_{+}$, particularly in the case of dominant attractive intra-band
pairing \cite{Efremov2011}. In Ref. \cite{Trevisan_PRL_2018}, Trevisan \etal
argued that an $s_{-}$ to $s_{+}$ change can also be induced
for a fixed disorder potential by changing the carrier concentration,
once the Lifshitz transition is crossed. In general, across this change
from $s_{-}$ to $s_{+}$, one of the gaps vanishes, making the system
behave effectively as a single-band superconductor (except in the
more exotic case where a time-reversal symmetry-breaking state emerges
\cite{Stanev14,Babaev17,Babaev18}). Furthermore, for large enough inter-band scattering,
the two gaps tend to the same value \cite{Golubov1997,Efremov2011}.
This effect was invoked by Thiemann \etal to explain why the behavior of the optical
conductivity of Nb-doped STO resembles that of single-band superconductors
\cite{Thiemann2018}.

\section{Perspectives}
\label{Sec:Perspectives}

The topics discussed above are but a few among various interesting
issues related to the superconducting properties of STO. Before finishing
this review, we briefly mention other interesting topics that,
in our view, also warrant further investigation.
\begin{itemize}

\item \textit{Relationship with LAO/STO. }The discovery of gate-tunable superconductivity
in LaAlO$_{3}$/SrTiO$_{3}$ (LAO/STO) interfaces and heterostructures
opened a new route to study superconductivity in the 2D limit \cite{Reyren2007}
(for a recent review, see Ref. \cite{Levy_review_2018}). An interesting
question is whether the superconducting state in LAO/STO is related
to that of STO, or whether it is a property of
the 2D electron-gas formed at the interface \cite{Millis2004}.
The existence of a $T_{c}(n)$ dome in LAO/STO, with a maximum $T_{c}$
value similar to that of STO, suggests that these phenomena are
related. Valentinis \etal proposed that the superconducting dome of LAO/STO
can be well modeled assuming that the pairing interaction of STO is
subject to quantum confinement \cite{Valentinis2017} (see also Refs.
\cite{Bianconi1994,Innocenti2010,Bianconi2014}). Within this perspective, LAO/STO would offer
another route to elucidate superconductivity in STO. An interesting
similarity in their phase diagrams is that the superconducting domes
of both LAO/STO and STO display a suppression of $T_{c}$ as a Lifshitz
transition is crossed \cite{Joshua2012,Singh2019}. Despite the similarities,
there are important differences: the explicit breaking of inversion
symmetry at the interface in LAO/STO leads to a different order of
bands, and to the removal of the spin-degeneracy of each band. 
Interestingly,
it has been recently proposed that this effect may result in an enhancement
of $T_{c}$ \cite{Haraldsen2012} or in a topologically non-trivial
superconducting state in LAO/STO \cite{Scheurer2015}. 

\item \textit{BEC-BCS crossover. }The fact that superconductivity in STO survives
down to very small carrier concentrations, corresponding to doping
levels below $0.01\%$ and Fermi energies of a few meV, raises the
question of whether the superconducting properties of STO could be described in terms of a BEC-BCS crossover. Indeed,
an early work by Eagles suggested that the formation of non-coherent Cooper pairs could onset even above the
superconducting transition temperature in Zr-doped STO \cite{Eagles1969}.
Interestingly, pre-formed pairs were recently reported in certain
LAO/STO heterostructures by Cheng \etal \cite{Levy2015} and pseudogap behavior was observed~\cite{richter2013interface}; 
whether these observations imply a BEC behavior remains under debate \cite{Hofman2017}. An important consideration
is that while in two dimensions the BEC behavior can appear already
for a weak attractive pairing interaction, a strong interaction is
needed in three dimensions, as relevant for STO \cite{Eagles1969}.
Moreover, the BEC-BCS crossover can be rather different in multi-band
systems \cite{Chubukov2016,Perali2019}. In this regard, we note that the superconducting
properties of STO are not the ideal ones for BEC behavior to be observed.
For the entire phase diagram, the zero-temperature gap $\Delta\sim T_{c}$
remains much smaller than the Fermi energy $\e_{F}$, even when the
latter is very small. Furthermore, the superconducting coherence length
is of the order of $100$ nm \cite{Collignon2017}, which would imply
the overlap of many Cooper pairs instead of
the tightly bound pairs expected for BEC behavior -- as pointed out by van der Marel \etal \cite{vanderMarel2011}. Finally, there are no signatures of pairing above $T_c$, such as strange metallic behavior~\cite{Lin20133}.

\item \textit{Antiferrodistortive domain walls. }A question that remains unsettled
is which impact, if any, the antiferrodistortive cubic-to-tetragonal transition that STO
undergoes at approximately $105$ K has on superconductivity [see Fig. \ref{fig:FE_diag}(c)].
Lin \etal proposed that filamentary
superconductivity originating from domain walls separating different
tetragonal domains is the reason why the superconducting
transition as marked by the onset of zero resistivity was observed
above the bulk $T_{c}$ of optimally Nb-doped STO \cite{Lin_nodeless_2015}.
Recent experiments in thin films of Nb-doped STO using a scanning
SQUID susceptometer by Noad \etal revealed a local enhancement of $T_{c}$ of about
$10\%$ as compared to the bulk $T_{c}$ \cite{Moler2016}. These
experiments, however, seem to favor a scenario in which the enhancement
of $T_{c}$ happens inside the tetragonal domains, rather than at
their boundaries. These STO domains were also observed by Wissberg \etal to modulate
the superconducting properties of films of different types of superconductors
grown on STO \cite{Kalisky2017}. Elucidating which of the several
local properties (electronic, dielectric, ferroelectric, etc) that are changed
inside the domains or at the domain walls correlate with the enhancement
of $T_{c}$ is therefore an important step to understand superconductivity
in STO. In this context, we point out the recent results of Pelc \etal correlating intrinsic structural inhomogeneity to the unusual temperature dependence of the superconducting fluctuations of STO, as measured by nonlinear magnetic response \cite{Pelc2019}. It would be interesting to establish whether this inhomogeneity is related to the AFD transition.

\item \textit{Normal-state transport properties. }Many unconventional superconductors,
such as iron pnictides, cuprates and heavy fermions display unusual
normal-state transport properties, chiefly manifested by a linear-in-$T$
resistivity, which contradicts the expectation of Fermi liquid theory.
At first sight, STO may seem to fall outside this category, since
its normal-state resistivity shows $T^{2}$ behavior at low temperatures
\cite{vanderMarel2011,lin2015scalable,Mikheev2016}, which is the standard
power-law expected from electron-electron scattering. The problem
is that Fermi liquid theory predicts that the resistivity of dilute
STO should not display $T^{2}$ behavior \cite{Walle2017}, one of
the reasons being the fact that, due to the smallness of the  Fermi
surface surrounding the $\G$-point [see Fig.~\ref{fig:TB-model}(b)], umklapp scattering is ineffective in relaxing momentum. Moreover,
the $T^{2}$ behavior is extended to temperatures higher than the
Fermi temperature \cite{lin2015scalable}. This does not mean that
the normal state of STO is not a Fermi liquid. Quite on the contrary, as shown in Fig. \ref{fig:sommerfeld},
specific heat measurements performed by McCalla \etal over a wide doping range found
an excellent agreement between the measured electronic Sommerfeld
coefficient and the predictions from a tight-binding model fitted
to DFT, provided that the effective mass is renormalized by a factor
of $2$ \cite{McCalla2019}. This renormalization factor, indicative
of a weakly-correlated system, was found to be independent of doping
(see also Ref. \cite{vanderMarel2011}). Such a disconnect between thermodynamic
properties, which suggest a standard Fermi liquid, and transport properties,
which suggest a ``non-Fermi liquid'' mechanism for $T^{2}$ resistivity,
is also manifested by the fact that STO does not follow the usual
Kadawoki-Woods scaling between the Sommerfeld coefficient and the
$T^{2}$ coefficient of the resistivity \cite{McCalla2019}. As shown by Lin \textit{et al.}, however, the $T^2$ coefficient does scale with $1/\e_F$ \cite{lin2015scalable}. Not
only is the origin of the $T^{2}$ resistivity in STO unsettled, but
also what relevance it has, if any, to the superconductivity of STO.
\end{itemize}

Thus, even after more than five decades since its discovery, superconductivity
in STO remains a challenging problem in which several contemporary
concepts in quantum matter research emerge. The resulting complex
landscape of electronic phenomena include: proximity to a putative
quantum critical point -- in this case, a little studied ferroelectric
metallic quantum phase transition; multi-band superconductivity beyond
the two-gap regime; pairing in the extreme dilute regime; unusual
normal-state transport properties. In this regard, by applying in
STO the powerful experimental and theoretical techniques developed
recently in the studies of other quantum materials, one has a promising
model system to potentially elucidate the connection between these
remarkable features, common to several quantum materials of interest,
and superconductivity.

\begin{acknowledgments}
We thank A. Balatsky, P. Barone, A. Bhattacharya, K.
Behnia, A. Chubukov, M. Greven, J. Haraldsen, B. Jalan, P. Lee, C.
Leighton, G. Lonzarich, J. Lorenzana, D. Maslov, V. Pribiag, B. Shklovskii, T. Trevisan, and P. W\"olfle for fruitful discussions. We thank Y. Ayino and T. Trevisan for their assistance in
producing figures \ref{fig_Hc2} and \ref{fig_disorder}, \ref{fig_Lifshitz_theory}, respectively. RMF and MNG were supported by the U. S.
Department of Energy through the University of Minnesota Center for
Quantum Materials, under Award No. DE-SC-0016371. 
During the writing of this manuscript, MNG was supported by the Italian MIUR through Project No. PRIN 2017Z8TS5B, and by Regione Lazio (L. R. 13/08) under project SIMAP.
JR acknowledges the support of the Israeli Science Foundation under grant No. 967/19.

\end{acknowledgments}

\bibliography{STO}

\end{document}